\documentclass[twocolumn]{aastex62}

\usepackage{graphicx}	
\usepackage{amsmath}	
\usepackage{amssymb}	
\usepackage{multirow}
\usepackage{color}
\usepackage{natbib}
\usepackage{morefloats}

\usepackage[T1]{fontenc}
\usepackage{ae,aecompl}

\begin{document}

\title{NEPTUNE'S 5:2 RESONANCE IN THE KUIPER BELT}
\shorttitle{Neptune's 5:2 Resonance}

\author[0000-0002-1226-3305]{Renu Malhotra}
\correspondingauthor{Renu Malhotra}
\email{renu@lpl.arizona.edu}
\affil{{Lunar and Planetary Laboratory, The University of Arizona, 1629 E University Blvd, Tucson, AZ 85721}}
\author{Lei Lan}
\affil{{Lunar and Planetary Laboratory, The University of Arizona, 1629 E University Blvd, Tucson, AZ 85721}}
\affil{{School of Aerospace Engineering, Tsinghua University, Beijing, 100084, China}}\author[0000-0001-8736-236X]{Kathryn Volk}
\affil{{Lunar and Planetary Laboratory, The University of Arizona, 1629 E University Blvd, Tucson, AZ 85721}}
\author{Xianyu Wang}
\affil{{School of Aerospace Engineering, Tsinghua University, Beijing, 100084, China}}

\shortauthors{Malhotra, Lan, et al.}

\begin{abstract}
{
Observations of Kuiper belt objects (KBOs) in Neptune's 5:2 resonance present two puzzles: this third order resonance hosts a surprisingly large population, comparable to the prominent populations of Plutinos and Twotinos in the first order 3:2 and 2:1 resonances, respectively; secondly, their eccentricities are concentrated near $0.4$. To shed light on these puzzles, we investigate the phase space near this resonance with use of Poincar\'e sections of the circular planar restricted three body model. We find several transitions in the phase space structure with increasing eccentricity, which we explain with the properties of the resonant orbit relative to Neptune's. The resonance width is narrow for very small eccentricities, but widens dramatically for $e\gtrsim0.2$, reaching a maximum near $e\approx0.4$, where it is similar to the maximum widths of the 2:1 and 3:2 resonances. We confirm these results with N-body numerical simulations, including the effects of all four giant planets and a wide range of orbital inclinations of the KBOs. {{} We find that the boundaries of the stable resonance zone are not strongly sensitive to inclination and remain very similar to those found with the simplified three body model, with the caveat that orbits of eccentricity above $\sim0.53$ are unstable; higher eccentricity orbits are phase-protected from destabilizing encounters with Neptune but not with Uranus. These results show that the 5:2 resonant KBOs are not more puzzling than the Plutinos and Twotinos; however, detailed understanding of the origins of eccentric, inclined resonant KBOs remains a challenge.}
}
\end{abstract}




\section{Introduction}

A striking feature in the orbital period distribution of Kuiper belt objects (KBOs) is the abundance of resonant orbits, that is, orbital periods of low integer ratios with Neptune's orbital period.  These resonant populations are of much interest for many reasons, particularly for the constraints that they can provide on the orbital migration history of Neptune and of the other giant planets \citep[e.g.][]{Malhotra:1995,Yu:1999,Chiang:2002,Murray-Clay:2005,Levison:2008,Kaib:2016}.  Many different resonances have been found to be occupied with KBOs \citep{Elliot:2005,Gladman:2012,Volk:2016}, but the most abundant populations appear to be in the 1:1, 3:2, 2:1, and 5:2 exterior mean motion resonances (MMRs) of Neptune (listed in order of decreasing mean motion).  Although the potential for high occupation of the 1:1, 3:2 and the 2:1 MMRs was anticipated in theoretical works~\citep[e.g.][]{Holman:1993,Malhotra:1995}, \cite{Chiang:2003} first noted that the high abundance of objects in the 5:2 MMR was particularly surprising and difficult to explain with proposed theoretical models.  The 5:2 MMR is a third order resonance (located at a semi-major axis of about 55 AU) and has larger separation from Neptune than the first order 2:1 and 3:2 MMRs, therefore it is expected to be relatively weak. 
The observational evidence that the 5:2 MMR hosts a large population of KBOs is rather astonishing.

After accounting for the observational biases, a recent analysis by \cite{Volk:2016} concludes that the total population in the 5:2 MMR is at least $\sim8500^{+7500}_{-4700}$ (for objects of absolute magnitude $H_r<8.666$, corresponding to 50--100 km in size); the authors note that this is similar to their estimated populations for the 2:1 and the 3:2 MMRs. 
(A previous estimate of the total population of the 5:2 MMR based on the results of the the Deep Ecliptic Survey \citep{Elliot:2005} found that it is slightly less populated than the 2:1 MMR and about half as populated as the 3:2 MMR \citep{Adams:2014}.)
Additionally, both \cite{Chiang:2003} and \cite{Volk:2016} remark on the orbital eccentricities of the 5:2 MMR KBOs which appear to be concentrated near eccentricity $e\approx0.4$.   

This twin puzzle -- the surprisingly large population of 5:2 resonant KBOs and the concentration of their eccentricities near 0.4  -- motivates us to investigate in more detail the phase space structure of this resonance.  We map the boundaries of the stable libration zone over the full range of eccentricity ($\sim0$ to $\sim1$) in the $(a,e)$ plane so as to facilitate comparisons with the widths of the 2:1 and 3:2 MMRs, and to provide dynamical context for the relative populations observed in these resonances.  We make use of non-perturbative numerical analysis of both the simplified model of the circular planar restricted three body problem of the Sun-Neptune-test-particle as well as three-dimensional N-body numerical simulations of test particle KBOs perturbed by the four giant planets.

This paper is organized as follows.  Poincar\'e surfaces of section near Neptune's 5:2 exterior MMR of the circular planar restricted three body model for a wide range of test particle eccentricities are presented and analyzed in Section 2.  Numerical N-body simulations of the more complete model of the Sun, the four giant planets and test particle KBOs are presented in Section 3.  In Section 4, we discuss the results in the context of the observational sample of the resonant KBOs.  We summarize and conclude in Section 5.

\section[]{Resonance width in the  Planar Circular Restricted Three Body Model}\label{s:pcr3bd}

A basic picture of the resonance phase space was outlined by \cite{Chiang:2003} who carried out numerical integrations of test particle trajectories for $3\times10^5$ years in the circular planar restricted three body model of the Sun-Neptune-test-particle. Their map of the initial conditions (in the semi-major axis $a$, and eccentricity $e$ parameter space) of librating, circulating and unstable test particle orbits gives a first approximate indication of the stable resonance libration width as a function of eccentricity, showing that the resonance widens considerably at eccentricities above $e\sim0.2$ (see their Figure 8); their study was limited to $e<0.6$.  Here we use Poincar\'e surfaces of section to more accurately visualize the extent of the stable resonant libration zone in the $(a,e)$ plane, over the full range of eccentricity.  Poincar\'e sections have been used previously by several authors to study Neptune's exterior resonances \citep[e.g.][]{Malhotra:1996,Hadjifotinou:2002,Kotoulas:2004,Voyatzis:2005b,Celletti:2007}.  In particular, \cite{Malhotra:1996} used Poincar\'e sections to map the boundaries in the $(a,e)$ plane of the stable libration zones of a few of Neptune's exterior resonances, and \cite{Wang:2017} used a similar approach to map the stable zones of the 2:1 and 3:2 interior resonances of the hypothetical planet 9.

\subsection{Methodology}\label{s:pcr3bd-methodology}

We follow the method described in \cite{Wang:2017} to compute the Poincar\'e sections, and from these we measure the resonance widths; we briefly describe this method here.  We adopt the circular planar restricted three body model of the Sun, Neptune and test particle (the latter representing a Kuiper belt object).  In this approximation, all the bodies move in a common plane, Neptune revolves around the Sun in a circular orbit, and the test particle revolves in an (osculating) elliptical heliocentric orbit. The masses of Sun and Neptune are denoted by $m_{\rm{1}}$  and $m_{\rm{2}}$, respectively. The fractional mass ratio of Neptune, $\mu=m_{\rm{2}}/(m_{\rm{1}}+m_{\rm{2}})= 5.146 \times 10^{-5}$, is very small. The third body, the massless test particle, has no influence on the motion of $m_{\rm{1}}$ and $m_{\rm{2}}$. We adopt the natural units for this model: the unit of length is the constant orbital separation of $m_1$ and $m_2$, the unit of time is their orbital period divided by $2\pi$, and the unit of mass is $m_1+m_2$; with these units the constant of gravitation is unity, and the orbital angular velocity of $m_1$ and $m_2$ about their common center of mass is also unity.
Then, in a rotating reference frame, of constant unit angular velocity and origin at the barycenter of the two primaries, both $m_1$ and $m_2$ remain at fixed positions, $ (-\mu,0) $ and $ (1-\mu,0) $, respectively, and we denote with ($x, y$)  the position of the test particle. The distances between the test particle and the two primaries can be written as
\begin{equation}\label{r1r2}
\begin{cases}
 r_1=\left[ (x+\mu)^2+y^2 \right]^{1/2}\\
 r_2=\left[ (x-1+\mu)^2+y^2 \right]^{1/2},
\end{cases}
\end{equation}	
and the equations of motion of the test particle can be written as,
\begin{equation}\label{xdotydot}
\begin{cases}
\ddot{x}=2\dot{y}+x-\frac{(1-\mu)(x+\mu)}{{r_1}^3}-\frac{\mu(x-1+\mu)}{{r_2}^3},
\\
\ddot{y}=-2\dot{x}+y-\frac{(1-\mu)y}{{r_1}^3}-\frac{\mu y}{{r_2}^3}.
\end{cases}
\end{equation}
where $``\cdot"$ and $``\cdot\cdot"$ represent the first and second derivative with respect to time.
These equations admit a conserved quantity, the Jacobi constant, $J$, given by,
\begin{equation}\label{jacobi}
J=x^2+y^2-\dot{x}^2-\dot{y}^2+\frac{2(1-\mu)}{r_1}+\frac{2\mu}{r_2},
\end{equation}
which can also be expressed in terms of the Keplerian orbital elements, $ a $ and $ e $,  the semi-major axis and eccentricity, respectively, of the particle's osculating orbit about the sun,
\begin{equation}\label{jacobiae}
J = \frac{1}{a}+2\sqrt{a(1-e^2)(1-\mu)}+O(\mu).
\end{equation}

\begin{figure}
 \centering
 \includegraphics[width=70mm]{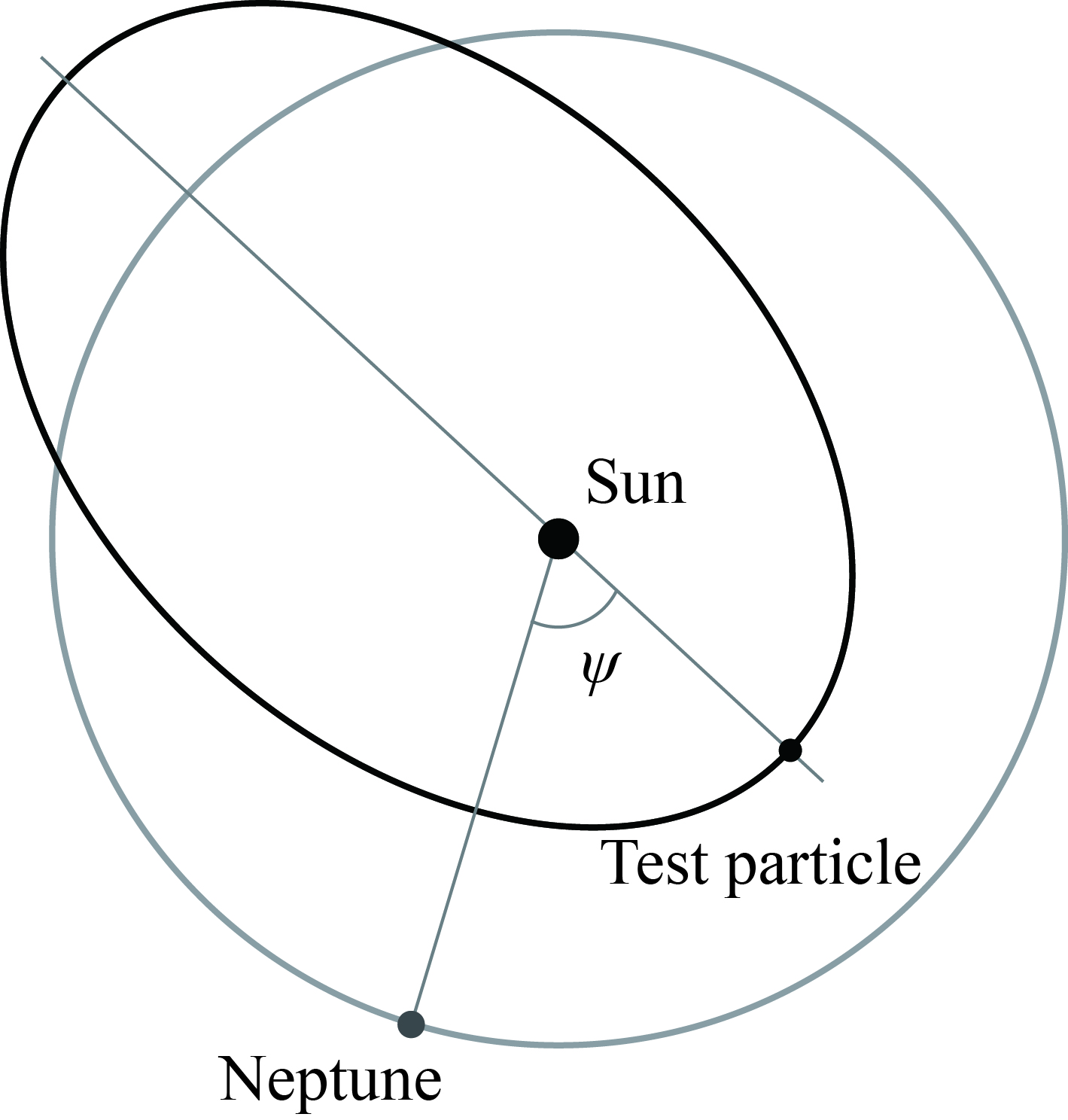}
 \caption{A schematic diagram to illustrate the definition of $\psi$, the angular separation of Neptune from the test particle at pericenter. }
 \label{f:f1}
\end{figure}

The Jacobi constant constrains a particle's motion to a three dimensional surface in its four dimensional phase space $(x,y,\dot{x},\dot{y})$.  To conveniently and meaningfully visualize in a two dimensional space the behavior of particles having the same Jacobi constant (representative of the 5:2 MMR, see below), we use the Poincar\'e surface-of-section technique. We define the surface of section by the condition  $\dot{r}_1=0,\ddot{r}_1>0$, as in \cite{Wang:2017}. That is, we record the state vector, $(x,y,\dot{x},\dot{y})$, of the particle at every perihelion passage.  We transform the test particle's position and velocity into osculating heliocentric orbital elements. Then we make the Poincar\'e sections as plots of $(a,\psi)$, where $\psi$ is the angular separation between the test particle and Neptune when the particle is at perihelion; the definition of $\psi$ is illustrated in Figure~\ref{f:f1}.

The equations of motion, Eq.~\ref{xdotydot}, are numerically integrated with the adaptive step size 7th order Runge-Kutta method~\citep{Fehlberg:1968} to obtain the continuous track of the test particle for 5000 Neptune orbital periods, starting from specified initial conditions. The relative and absolute error tolerances are controlled to be lower than $10^{-12}$.  For each Poincar\'e section, we specify the Jacobi constant and choose several different initial conditions, all near the 5:2 MMR as follows.  The initial $a$ is always chosen as the resonant orbital semi-major axis, $a_{\rm{res}}$, which can be calculated from Kepler's third law:
\begin{equation}\label{ares}
a_{\rm{res}}=(\frac{5}{2})^{2/3} \simeq 1.8420.
\end{equation}
Then by specifying the initial eccentricity, $e_i$, we specify the Jacobi constant by means of Eq.~\ref{jacobiae}.
For different particle trajectories having the same Jacobi constant (i.e., for the same Poincar\'e section), we choose different initial values of $\psi$.  These initial conditions completely define the initial position and velocity of the particle:
\begin{align}\label{xpyp}
(x_{\rm{p}},y_{\rm{p}}) &= r_{\rm{p}}(-\mu+\cos\psi,-\sin\psi), \nonumber \\
(\dot{x}_{\rm{p}},\dot{y}_{\rm{p}}) &=  v_{\rm{p}}(\sin\psi,\cos\psi),
\end{align}
where $r_{\rm{p}}=a_{\rm{res}}(1-e_i)$ is the initial perihelion distance, and $v_{\rm{p}}$ is the norm of the particle's velocity vector at perihelion; $v_{\rm{p}}$ is computed from Eq.~\ref{jacobi} as follows,
\begin{equation}
v_{\rm{p}}=\sqrt{ x_{\rm{p}}^2+y_{\rm{p}}^2+\frac{2(1-\mu)}{r_1}+\frac{2\mu}{r_2}-J}.
\end{equation}

For future reference, we introduce the usual ``critical resonant argument" defined by
\begin{equation}\label{e:phi}
\phi = 5\lambda - 2\lambda_N-3\varpi,
\end{equation}
where $\lambda,\lambda_N$ are the mean longitudes of the particle and of Neptune, respectively, and $\varpi$ is the particle's longitude of perihelion. For every point on the Poincar\'e section, the particle is located at its perihelion (that is, $\lambda=\varpi$), and $\psi=\varpi-\lambda_N$. Therefore, we have the following relationship between $\phi$ and $\psi$:
\begin{equation}\label{e:phi-psi}
\phi = 2(\varpi - \lambda_N) = 2\psi.
\end{equation}

On the $(a,\psi)$ Poincar\'e section, the exact 5:2 stable resonant orbit (which is a periodic orbit of the restricted three body model) appears as a pair of points at the center of the resonance zone; the test particle with this orbit intersects the surface of section alternately at each of this pair of points. The two points have the same value of semi-major axis but different values of $\psi$, differing by 180$^\circ$. (They have the same value of the critical resonant argument, $\phi$.) The entire resonance zone is located in the neighborhood of $a_{\rm{res}}$, and it appears as ``islands" in the Poincar\'e section surrounding each of the pair of points of the exact stable resonant orbit. In these resonance islands, stable quasi-periodic orbits librate about the exact resonant orbit, characterized by a libration of $\psi$ over a limited range of values; this corresponds to the test particle orbit intersecting with the surface of section at points that lie on smooth closed curves enclosing the central points.  The island is bounded by either a separatrix (beyond which $\psi$ circulates rather than librates) or -- more often -- by a chaotic zone in which $\psi$ (and $a$) both exhibit irregular behavior.  In the chaotic zone, the successive points of a test particle's trajectory do not remain confined to a smooth curve, rather they scatter around and, over time, fill an area on the $(a,\psi)$ plane.  {{} (Note that the particle eccentricity also exhibits corresponding variations, while the combination of $a$ and $e$ is preserved as dictated by the Jacobi constant.)} The chaotic zone is understood to result from the interaction of neighboring high-order and secondary resonances~\citep{Chirikov:1979}.

We measure the upper and lower boundaries of the stable resonance islands around $a_{\rm res}$, as illustrated in Figure~\ref{f:f2}(e). In some cases, numerous small islands can be found near the resonance boundary which may make the resonance boundary blurry and hard to confirm visually. In this situation, we assume that these small islands belong to the chaotic zone, and we take the boundaries of the stable libration zone as the one interior to these clusters of small islands. The resonance width {{} in semi-major axis}, $\Delta a$, is measured as the difference between the upper and lower boundaries of the stable islands.  

We note that, in general, the width of a resonant island in such Poincar\'e sections depends on the resonance (i.e., $a_{\rm{res}}$), the mass ratio $\mu$, and the Jacobi constant $J$.  For the problem of interest here, the values of $\mu$ and $a_{\rm{res}}$ are fixed.  With the use of Eq.~\ref{jacobiae}, we use $e_i$ as a proxy for $J$.  We compute Poincar\'e sections for many values of $e_i$ spanning its full range $\sim0$ to $\sim1$.  Then, with all the Poincar\'e sections in hand for the many values of $e_i$, we map the resonance zone boundaries in the $(a,e_i)$ plane. 

\subsection{Surfaces of section: variation with test particle eccentricity}\label{s:pcr3bd-sos}

We calculated the surfaces of section in the neighborhood of the 5:2 MMR (semi-major axes near $a_{\rm{res}}$, Eq.~\ref{ares}), for fixed mass ratio $\mu=5.146\times10^{-5}$. {{} As described above, each surface of section has a fixed value of the Jacobi constant. The many different particle trajectories computed for each surface of section are given the same initial values of the semi-major axis and eccentricity, but different initial values of $\psi$ (spanning the full range $0^\circ$ to $360^\circ$). The initial semi-major axis of all particle trajectories were set to the unperturbed resonant value, $a_i=a_{\rm res}$, where the latter is given by Eq.~\ref{ares}.  Because a surface of section has a fixed value of the Jacobi constant, this choice of initial semi-major axis, $a_i=a_{\rm res}$, for all particle trajectories means that all particle trajectories in a particular surface of section will have the same initial value of the eccentricity (from Eq.~\ref{jacobiae}); different surfaces of section will have different initial values of the eccentricity. Importantly, this initial eccentricity is very close to the value of the eccentricity at the center of the stable islands. We therefore label each surface of section with the specified initial value of the eccentricity, and we drop the subscript on $e_i$ for simplicity.} Several sample surfaces of section are shown in the left panels in Figure~\ref{f:f2}.  

Alongside each surface of section, we also plot (in the right panels {{} in Figure~\ref{f:f2}}) the trace of the particle's exact resonant orbit in the rotating frame, to illustrate the geometry of resonant orbits at the corresponding values of the particle eccentricity. In these panels, we deliberately do not explicitly indicate the orientation of the $x,y$ axes of the rotating frame.  Instead, we indicate the location of the Sun (always close to the origin of the reference frame) and a gray circle of radius $1-\mu$ centered at the origin of the reference frame. The reader must imagine the positive $x-$axis as the line oriented from the location of the Sun towards Neptune; Neptune's location is fixed in the rotating frame and is somewhere on the gray circle.  The dots on the gray circle indicate the (multiple) possible locations of Neptune for stable resonance geometries, whereas the crosses indicate the (multiple) possible locations of Neptune for unstable resonance geometries.  Note that the trace the test particle's resonant trajectory in the rotating frame has a two-fold symmetry about the Sun-Neptune line.

When the eccentricity is low, as in Figure~\ref{f:f2}(a), we observe that the surface of section has a single pair of stable islands containing smooth closed curves, and there is no visible chaotic region. The two stable islands are centered at $\psi=90^{\circ}$ and $\psi=270^{\circ}$, with $a=1.842$. For reasons that will become apparent below, we call these the first resonance zone or Zone-I. Their boundary is a separatrix that passes through a pair of unstable points located at $\psi=0^{\circ}$  and $\psi=180^{\circ}$. The centers of the resonant islands correspond to the exact resonant orbit which is stable, that is, the stable periodic orbit in the 5:2 exterior MMR with Neptune.  In the right panels in Figure~\ref{f:f2}, the black dots on the gray circle indicate the position of Neptune (in the rotating frame) corresponding to the stable geometry of the exact 5:2 resonant orbit, whereas the crosses on the circle indicate the position of Neptune corresponding to the unstable geometry of the exact 5:2 resonant orbit. The stable and unstable geometries appear as pairs of points in the surface of section. Other smooth curves which traverse this surface of section correspond to the orbits that either librate around the stable exact resonant orbit or circulate across the entire range of $\psi$ from $0^{\circ}$ to $360^{\circ}$. The librating orbits can be thought of as a slow libration of the trace of the test particle orbit shown in the right panel, Figure~\ref{f:f2}(a), while the Sun and Neptune remain at their fixed locations in the rotating frame.

\begin{figure*}
\centering
 \includegraphics[width=174mm]{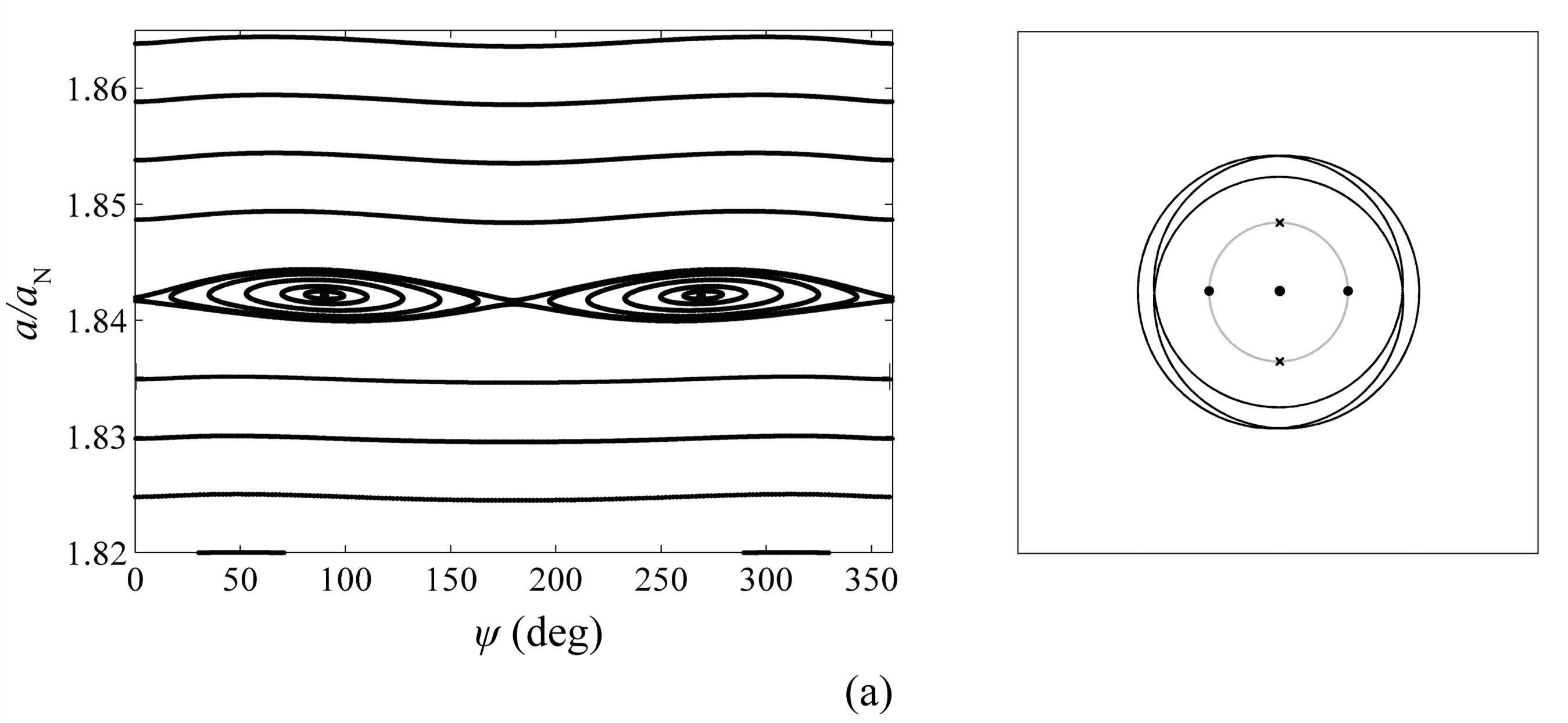}
\end{figure*}

\begin{figure*}
\centering
 \includegraphics[width=174mm]{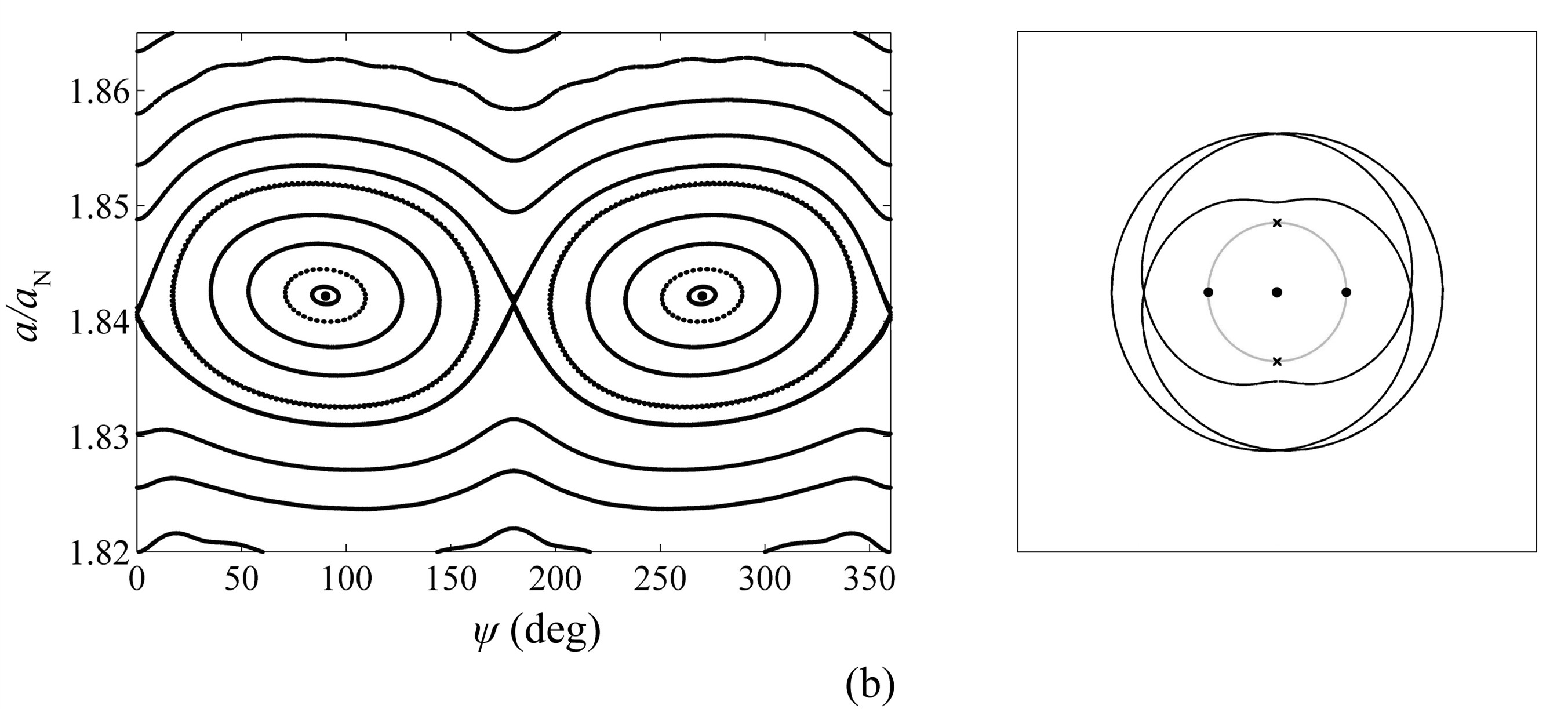}
\end{figure*}

\begin{figure*}
\centering
 \includegraphics[width=174mm]{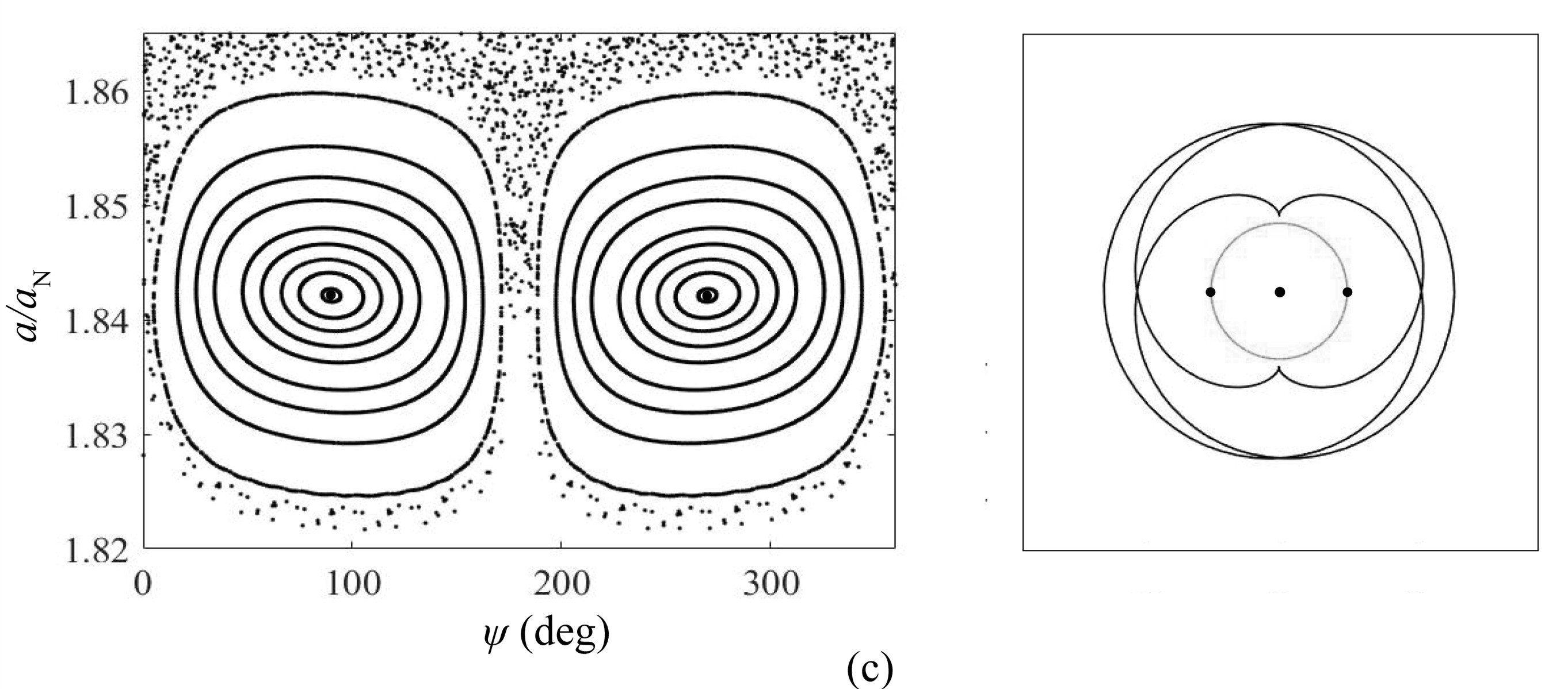}
\end{figure*}

\begin{figure*}
\centering
 \includegraphics[width=174mm]{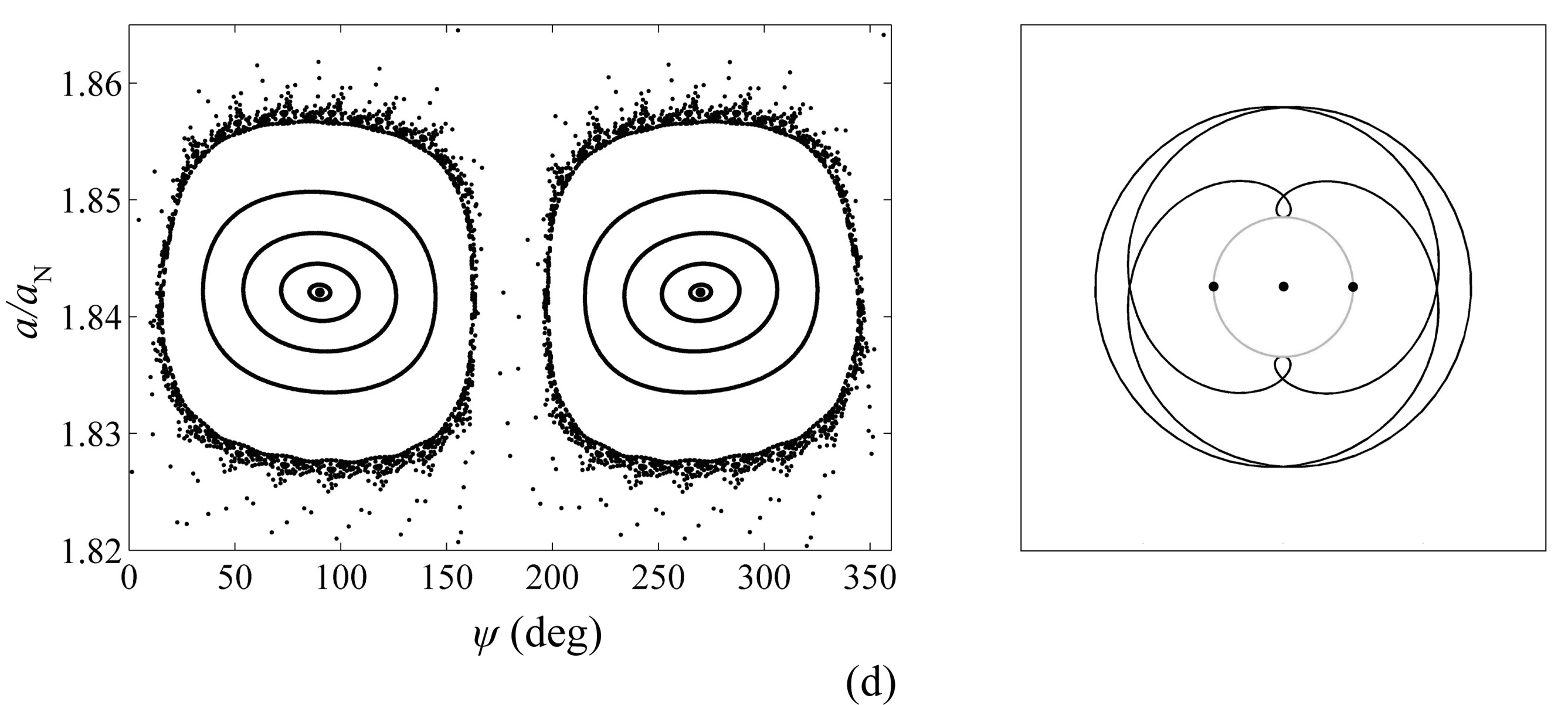}
\end{figure*}

\begin{figure*}
\centering
 \includegraphics[width=174mm]{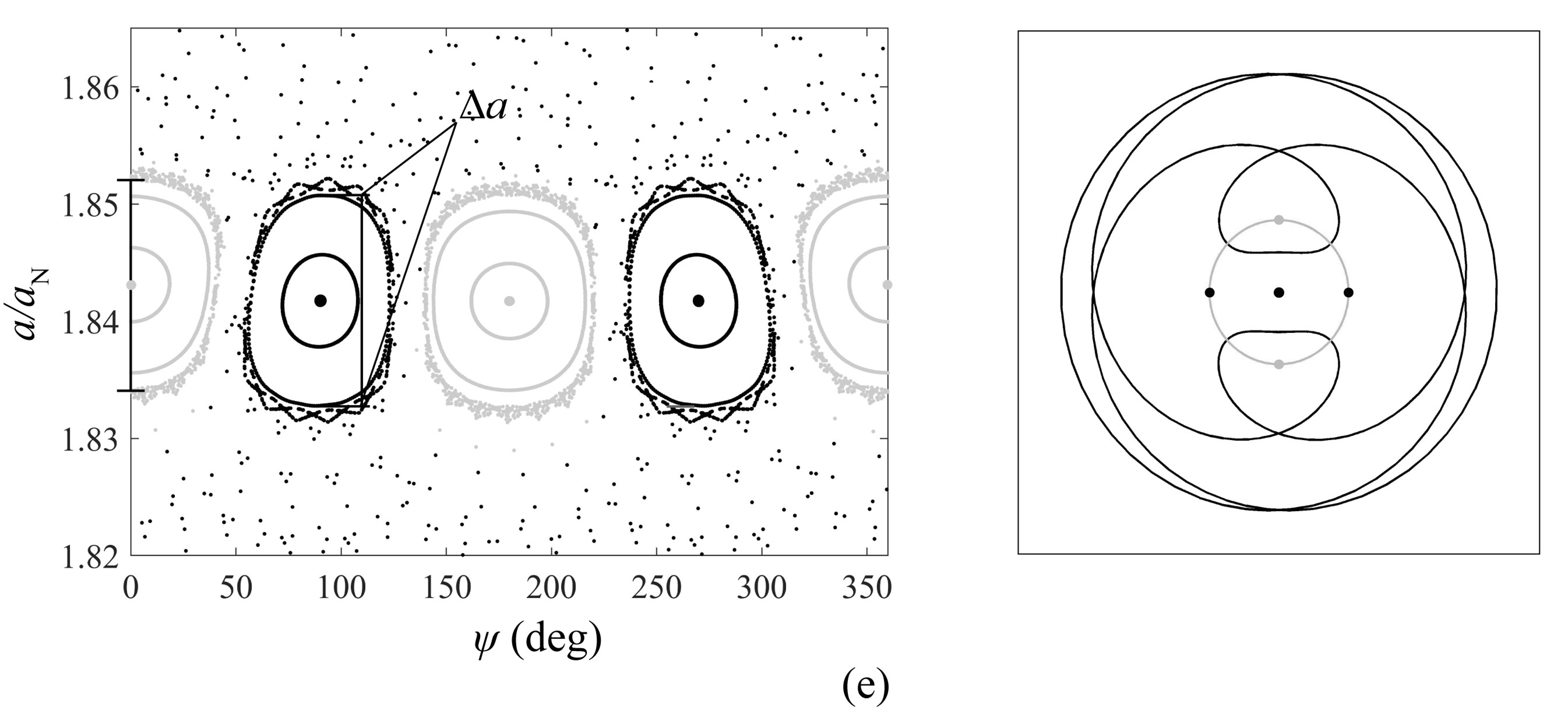}
\end{figure*}

\begin{figure*}
\centering
 \includegraphics[width=174mm]{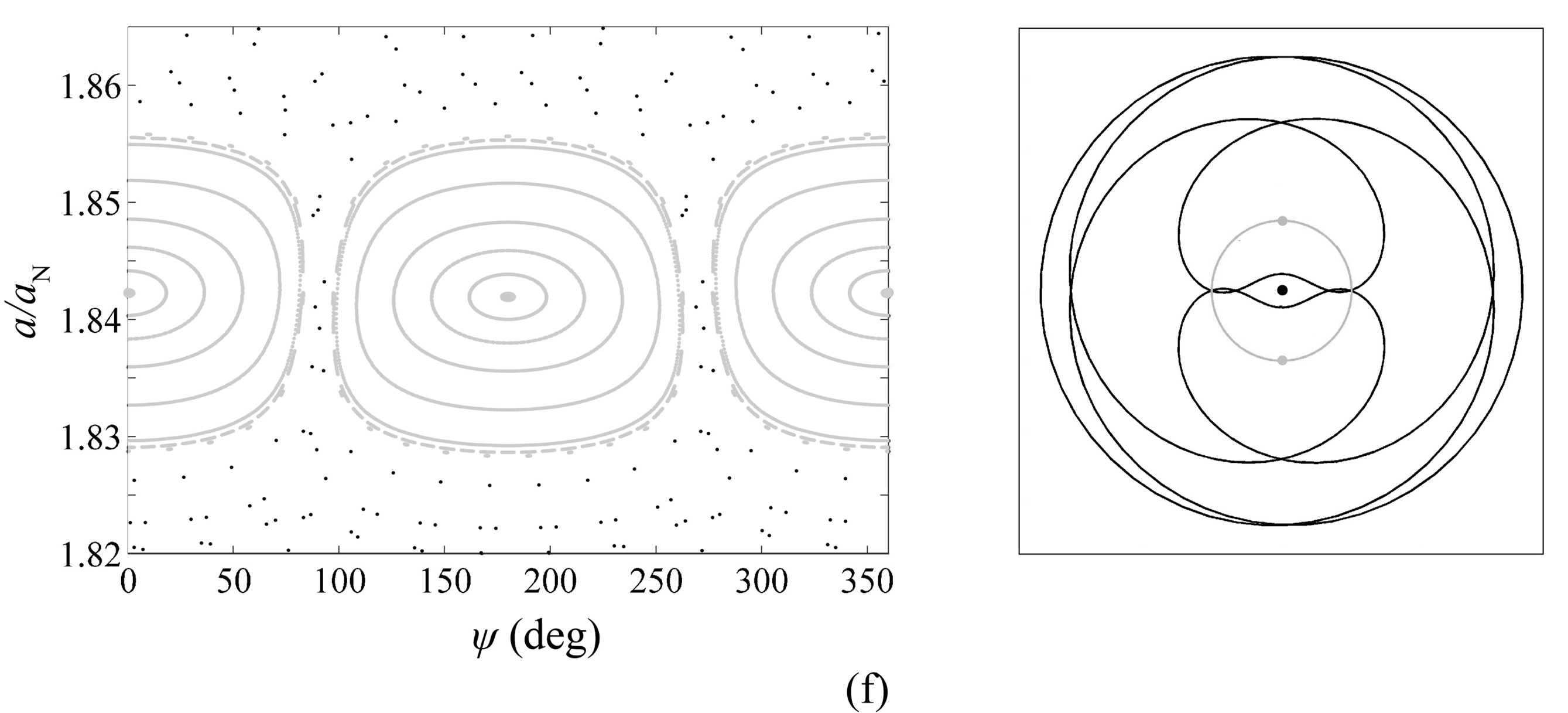}
\end{figure*}

\begin{figure*}
\centering
 \includegraphics[width=174mm]{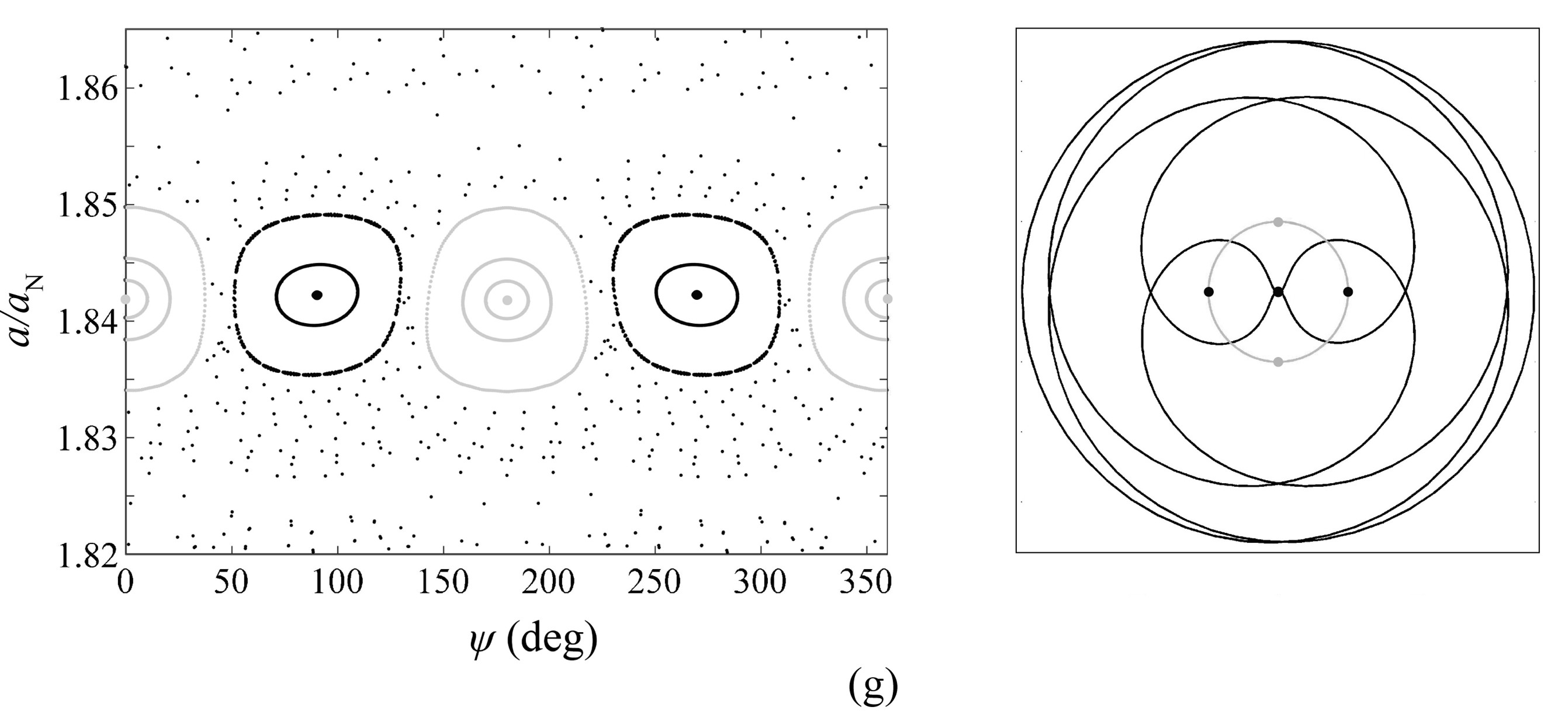}
\caption{(a--g) The surfaces of section (left panels) and the trace of the particle's resonant orbit in the rotating frame (right panels) at different {{} initial values of the particle eccentricity}: (a) $e=0.10$; (b) $e=0.30$; (c) $e=0.40$; (d) $e=0.457$; (e) $e=0.70$; (f) $e=0.87$; (g) $e=0.99$. In the left panels, the smooth curves in the libration islands shown in black dots are Zone I-1 or Zone I-2, and those shown in gray dots are Zone II. In the panels on the right, the gray circle is of radius $1-\mu$. The dots on this circle indicate Neptune's position corresponding to the geometry of stable centers of resonance islands; black and gray dots correspond to the centers of the Zone I-1 (or Zone 1-2) and Zone 2, respectively. The black crosses indicate Neptune's position corresponding to the geometry of the unstable points on the separatrix of Zone 1-1 (only at lower eccentricities when there is no visible chaotic zone). 
}
 \label{f:f2}
\end{figure*}

At larger eccentricity, the perihelion distance is lower and the widths of stable islands expand, as shown in Figure~\ref{f:f2}(b). For $e=0.30$, there is no visible chaotic area, and only smooth curves are seen in this surface of section.  But, for even larger eccentricity, $e=0.40$, as shown in Figure~\ref{f:f2}(c), the smooth separatrix curve disappears, and chaotic zones appear in the area between the two stable islands. The widths of two stable islands increase with eccentricity, reaching a maximum at eccentricity $e_{\rm{m1}}=0.402$.

We can understand this behavior as follows. At lower eccentricities, the test particle's trajectory lies entirely beyond Neptune's orbit, and the perihelion of the particle is far from the planet, thus the planet's perturbation is relatively weak for this planet-sun mass ratio. With the increase of eccentricity, the perihelion of the test particle's orbit reaches closer to Neptune, thus, the test particle is perturbed more strongly by the planet. When the test particle's eccentricity exceeds the Neptune-crossing value, $e_{\rm{c}}=1-a_{\rm{res}}^{-1}=0.457$, its orbit crosses Neptune's orbit, as shown in Figure~\ref{f:f2}(d).  At an eccentricity $e_{\rm{c1}}=0.473$, which slightly exceeds $e_{\rm{c}}$, a new pair of stable islands become visible in the surface of section, and these new islands are larger at even higher eccentricities. These new stable islands are centered at $\psi=0^{\circ}$ and $\psi=180^{\circ}$, in-between the original pair of stable islands found at lower eccentricities. We call this new pair of resonance islands ``Zone II".  The appearance of new stable islands indicates the presence of a new pair of periodic orbits in the 5:2 exterior MMR. 
Although these new and old periodic orbits trace the same shape in the rotating frame, {{} their orientations relative to the planet are distinctly different}.  
This is evident when we compare the different cases of Neptune's location in the right panel in Figure~\ref{f:f2}(e).  The possible positions of Neptune corresponding to the geometry of the resonance Zone-II are represented by gray dots on the gray circle in Figure~\ref{f:f2}(e).  {{} In other words, the new pair of stable islands represents a new pair of stable orientations of the particle's perihelion relative to Neptune's location.}

{{} Such doubling of stable islands has been attributed previously to a bifurcation of periodic orbits from a collision orbit \citep{Voyatzis:2005}, but here we have shown that it can be also explained physically by reference to the trace of the resonant orbit in the rotating frame.}  

For increasing eccentricity above $e_{\rm{c1}}$, the old islands of Zone I-1 shrink while those of Zone II expand at the expense of the old ones. Actually, once the perihelion distance of the test particle is lower than the radius of Neptune's orbit, as shown in the right panel of Figure~\ref{f:f2}(e), the two perihelion lobes of the particle's trajectory are cut by the  planet's circular orbit. We observe that the libration range of $\psi$ for the new stable islands is strongly related to the lengths of arcs of the planet's orbit which are interior to the perihelion lobes of the particle's orbit. The longer the length of these arcs, the wider the stable islands.

At an eccentricity of $\sim$0.70, as shown in Figure~\ref{f:f2}(e), the new and old stable islands have similar sizes, while the lengths of arcs of the planet's orbit inside and outside the perihelion lobes of the particle's orbit are similar as well. At higher eccentricities, the stable islands of Zone II increase in width, while those of Zone I-1 decrease. At eccentricity $e_{\rm{c2}}=0.872$, the old stable islands of Zone I-1 disappear. Almost at the same eccentricity, $e_{\rm{m2}}=0.874$, the widths of the stable islands of Zone II reach their maximum. Near this eccentricity, we observe that the arcs of the planet's orbit outside the perihelion lobes of the particle's orbit are crushed to vanishing length, as shown in Figure~\ref{f:f2}(f).

At even higher eccentricity, the two perihelion lobes of the particle's orbit overlap. At the critical value of eccentricity $e_{\rm{c3}}=0.909$, a pair of stable islands emerges at the same centers as the first Zone I-1 islands, i.e., centered at $\psi=90^{\circ}$  and $\psi=270^{\circ}$ (and $a=1.842$); we call these Zone I-2.  At eccentricity exceeding $e_{\rm{c3}}$,  the Zone I-2 islands grow in width along with the increase of the arc lengths interior to the overlapped regions of the test particle's perihelion lobes. Meanwhile, the Zone II islands shrink slightly.

At eccentricity $e$ values approaching $\sim$0.99, the perihelion distance of the particle is very close to the Sun. The arc lengths inside and outside the overlapped perihelion lobes are similar. Therefore, we observe that the Zone II islands are only a little larger in width than the Zone I-2 islands, as shown in Figure~\ref{f:f2}(g).

For the pair of resonant islands in Zone I-1 and I-2, the resonant angle, $\phi$, librates about a center at $180^\circ$, but for the pair of resonant islands in Zone II $\phi$ librates about $0^\circ$.

As an aside, we make note of an interesting property of the Zone II islands for eccentricities in the range $e_{\rm{c1}}=0.457$ to $e_{\rm{c2}}=0.872$. We observe that the two islands of Zone II are not centered at the same value of the semi-major axis. One is centered at $a=1.8417$ (at $\psi=180^\circ$), and the other is centered at $a=1.8431$ (at $\psi=0^\circ$). This difference is explained by the different positions of Neptune at the two successive perihelion passages of the test particle, which cause a slight difference in the perihelion distance and velocity and hence a slight difference in the osculating orbital parameters at alternate perihelion passages.  We can also note that, in this range of eccentricity, the geometry of the Zone II exact resonant orbit is symmetric about the $x$-axis but not about the $y$-axis in the rotating frame; at eccentricity exceeding $e_{\rm{c2}}$, the orbit becomes symmetric about the $y$-axis also.  The exact resonant orbits of Zone I-1 and Zone I-2 are symmetric about both the $x$-axis and the $y$-axis. {{} For the purpose of measuring the resonance width (see below in Section~\ref{s:pcr3bd-ae}), we use the Zone II island at $\psi=180^\circ$; the alternative choice makes only a minor difference in the measured resonance width.}

\subsection{Resonance zone in the $(a,e)$ plane}\label{s:pcr3bd-ae}

\begin{figure}
 \centering
 \includegraphics[width=250px]{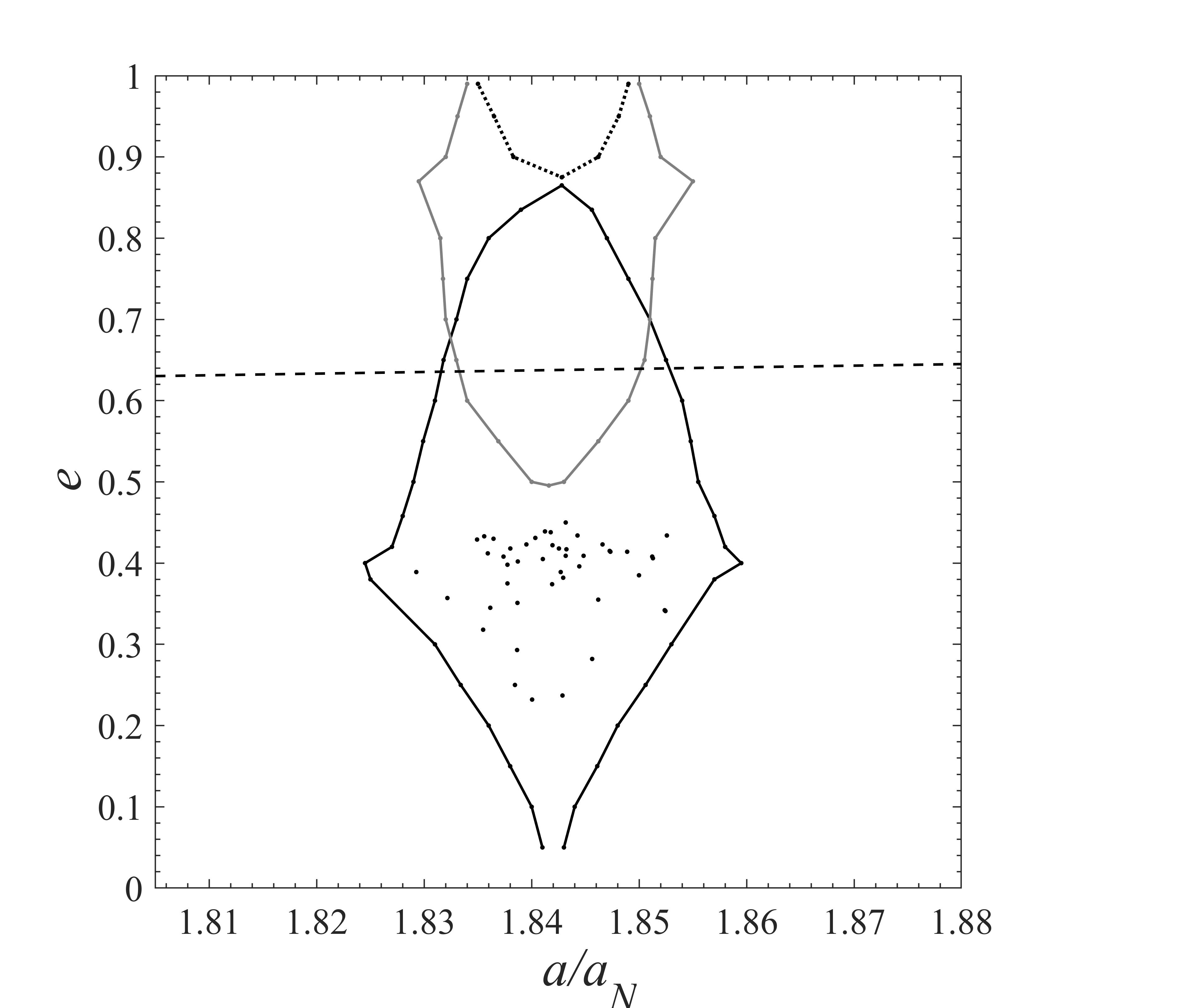}
\caption{The boundaries of the stable libration zones of Neptune's 5:2 exterior MMR in the $(a,e)$ plane; $a$ and $e$ are the osculating barycentric elements. {{} (As described in Section 2, the boundaries in semi-major axis are the minimum and maximum values of $a$ of the stable resonant islands in the Poincar\'e sections, while the eccentricity $e$ is that of the resonant orbit near the center of the stable island.)} The solid black curves indicate the boundaries of Zone I-1, the dotted black curves indicate those of Zone I-2, and the gray curves represent those of Zone II. The scattered black dots indicate {{} the osculating barycentric $a$ and $e$ of the observed Kuiper Belt objects identified to be} librating in Neptune's 5:2 MMR (see Section~\ref{s:observations}). The nearly horizontal dashed line indicates orbits of perihelion distance equal to Uranus' aphelion distance.  }
 \label{f:f3}
\end{figure}

By visually examining many Poincar\'e sections, we identified the boundaries, $a_{\rm{min}}$ and $a_{\rm{max}}$, of the stable resonance islands of Zone I-1, Zone II and Zone I-2, {{} as illustrated in Figure 2(e).  These boundaries are plotted in Figure~\ref{f:f3} as a function of eccentricity. (As noted previously, here eccentricity refers to the common initial eccentricity of test particle trajectories plotted in a Poincar\'e section; this value is a proxy for the Jacobi constant that characterizes a Poincar\'e section, and is also very close to the value of the resonant orbit at the center of the stable resonant islands.)}  We observe that the width, $\Delta a = a_{\rm max}-a_{\rm min}$, of Zone I increases at first and then decreases, finally disappearing; its largest extent occurs at eccentricity $e_{\rm{m1}}\simeq0.402$. The resonance islands of Zone II first appear at $e_{\rm{c1}}=0.473$; their widths increase at first and then decrease, and their largest extent occurs at $e_{\rm{m2}}\simeq0.874$.  The Zone I-2 resonance islands first appear at eccentricity $e_{\rm{c3}}\simeq0.909$ and increase in width as the eccentricity approaches 1.

Examining Figure~\ref{f:f3} further, it is noteworthy that at their largest extent, the widths in semi-major axis of the Zone II resonance islands are only modestly smaller than those of the Zone I-1 islands.
However, so far no stable resonant KBOs have been detected in Zone II (see Section~\ref{s:observations} and the Appendix).  This can be understood as follows.  For sufficiently large eccentricity, the perihelion distance of the test particle's orbit will drop below the orbit of Uranus and the particle is then likely to have gravitational scattering encounters with that planet which would remove it from Neptune's 5:2 resonance.  In Figure~\ref{f:f3}, we plot a dashed line to indicate the combination of the particle's $a$ and $e$ for which its perihelion distance would equal Uranus' aphelion distance; above this dashed line any resonant KBOs would be vulnerable, over long times, to instability arising from scattering encounters with Uranus.  We can then conjecture that in the region bounded by the gray curves but below the dashed line is where 5:2 resonant KBOs may potentially exist in Zone II, that is, with the resonant angle $\phi$ librating about $0^\circ$, in contrast with librations centered at $180^\circ$ in Zone 1-1.  However, numerical simulations with the full N-body model including the other giant planets' perturbations find that this region cannot host long term stable orbits (see Section 3 below).

\begin{figure}
 \centering
 \includegraphics[width=85mm]{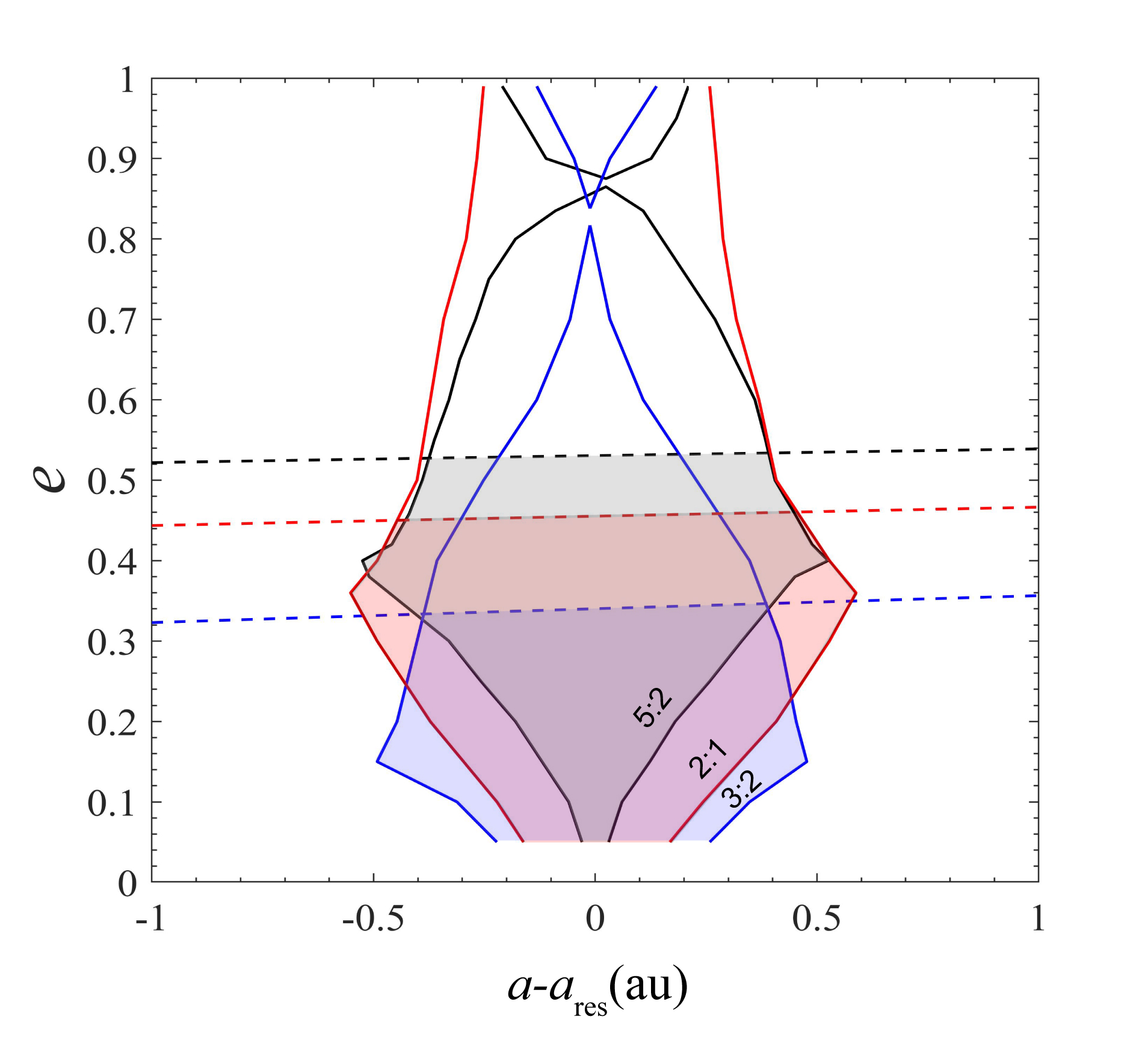}
\caption{{} The boundaries  in the $(a,e)$ plane of Neptune's 5:2 stable resonance Zone I-1 (in black) compared with those of the 3:2 MMR (in blue) and the 2:1 (in red).  (The latter two are from a forthcoming paper, Lan \& Malhotra, in preparation.)  The upper boundaries of the shaded zones are defined by perihelion distance equal to 26~au; we anticipate that resonant orbits with perihelion distance below this value remain phase-protected from destabilizing close encounters with Neptune but not from Uranus, on timescales of $10^7$~yr.}
 \label{f:MMRcompare}
\end{figure}

Translating the resonance width, $\Delta a$, to astronomical units, we find that at its largest extent the width of Zone I-1 is $\sim1.14$~au (near $e\simeq0.4$).
We can compare this width with the widths of Neptune's 3:2 and 2:1 MMRs.  In \cite{Malhotra:1996}'s study (see their Figure~11), the 3:2 and the 2:1 MMRs have maximum width of about 1 au, near $e\simeq0.15$ and near $e\simeq0.3$, respectively\footnote{We note the curious coincidence that the perihelion distance corresponding to the eccentricity of the maximum of the resonance width is near 33 au for all of these MMRs.}.  
{{} In a forthcoming paper, Lan \& Malhotra (in preparation) have calculated the boundaries of a large number of Neptune's exterior MMRs in the $(a,e)$ plane with higher resolution than in \cite{Malhotra:1996}'s study; in Figure~\ref{f:MMRcompare} we reproduce their results for the 3:2 and the 2:1 MMR in order to make a direct comparison with the 5:2 MMR. In this figure, we have also indicated an upper boundary at perihelion distance equal to $26$~au.  (We show in Section~\ref{s:numerical-sims} that this is the approximate boundary beyond which the 5:2 resonant KBOs are unstable on $10^7$~year timescales due to close encounters with Uranus; we conjecture that a similar perihelion distance boundary applies for long term stability in the 3:2 and the 2:1 MMRs.) We observe that while the shapes of these stable zones differ amongst the resonances, their overall sizes are not so different.  The area of Zone I-1 of the 5:2 MMR in the $(a,e)$ plane (integrated up to the stable boundary near $e\simeq0.53$) is about 84\% of the corresponding stable zone of the 3:2 MMR, and the area of the 2:1 stable zone is 1.005 times that of the 3:2 stable zone.}
In other words, the sizes of the stable libration zones of these three resonances are similar to each other.  

\begin{figure}
 \centering
 \includegraphics[width=85mm]{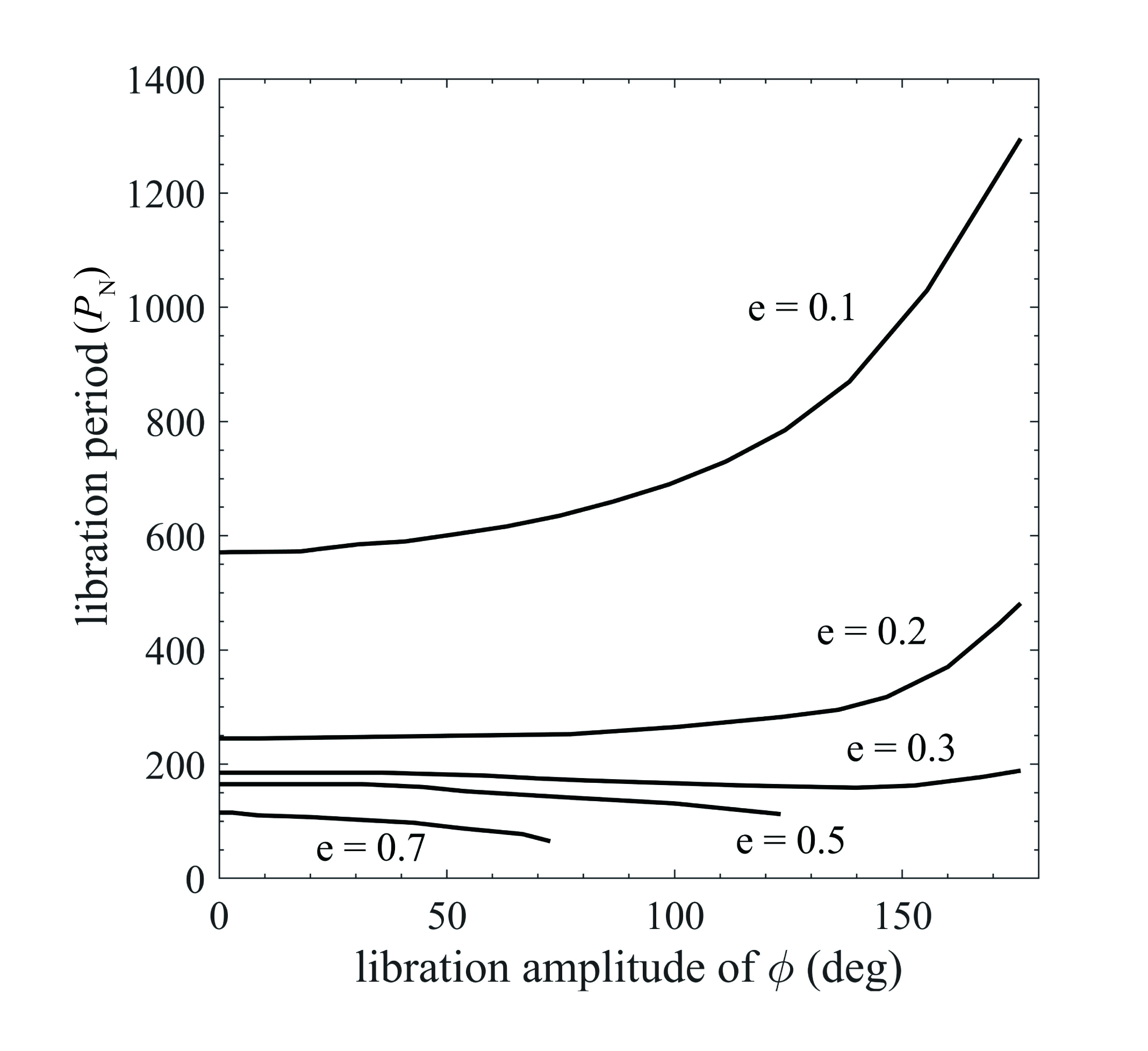}
\caption{{} Zone I-1 libration period (in units of Neptune's orbital period, $P_{\rm N}$) as a function of libration amplitude of resonant angle $\phi$, for representative values of the particle eccentricity (indicated).}
 \label{f:libamp}
\end{figure}

{{} For completeness of the description of the dynamics in the 5:2 MMR, we determined the libration periods as a function of the amplitude of libration of the resonant angle $\phi$ (Eq.~\ref{e:phi}) by examining the time series of the trajectories that we computed for the Poincar\'e sections of Figure~\ref{f:f2}. These are shown in Figure~\ref{f:libamp}, for several representative values of the eccentricity in Zone I-1. First, we observe that for librations of vanishing amplitude, the libration period decreases monotonically with increasing eccentricity, qualitatively similar to the pendulum model for resonance~\citep{Murray:1999SSD}, however, quantitatively, the eccentricity-dependence is not as simple as predicted by the pendulum model for resonance. Secondly, we observe that the trend with libration amplitude has two regimes of behavior. For smaller eccentricities, $e<0.3$, the libration period increases monotonically with libration amplitude, qualitatively similar to the pendulum model, but for larger eccentricities, the libration period decreases slightly with amplitude of libration.  The boundary of these opposite trends is near eccentricity $\sim0.3$, for which we observe that the behavior is not monotonic: the libration period decreases slightly as the amplitude of libration increases but reverses near a libration amplitude of $\sim150^\circ$ beyond which it increases. Also noteworthy is that for eccentricities above $e_{\rm c1}=0.473$, the maximum possible amplitude of libration of $\phi$ decreases sharply with increasing eccentricity as Zone II competes for phase space with Zone I-1; in this regime, in the limited range of libration amplitudes of $\phi$ the libration period is a monotonically decreasing function of amplitude.  In light of these results, the statement in \cite{Chiang:2003} that ``[in Neptune's 5:2 MMR] the libration period increases with decreasing libration amplitude, unlike the case for the conventional pendulum model for a resonance", holds for the parameter range of many of the presently known 5:2 resonators (because they are concentrated near eccentricities of $\sim0.4$), but is not generally accurate for all potential 5:2 resonators. This complex behavior of the libration period with eccentricity and with libration amplitude is common to several other Neptune's MMRs in the Kuiper belt~\citep{Malhotra:1996}.  A theoretical understanding of this behavior would be valuable, but is beyond the scope of the present work.}

\section{N-body numerical simulations}\label{s:numerical-sims}

\begin{figure}
 \centering
 \includegraphics[width=250px]{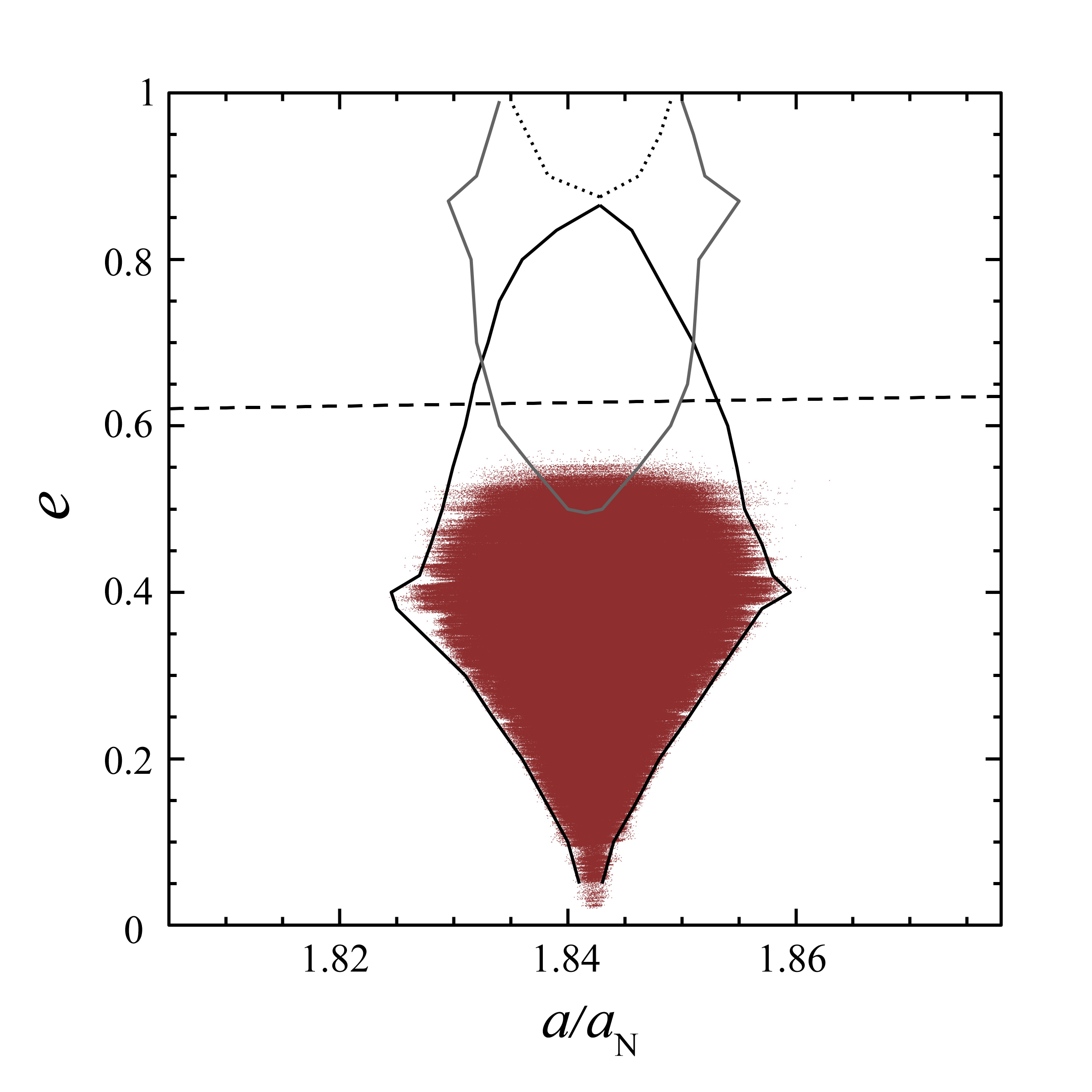}
\vglue-0.2truein
 \caption{Scatter plot of the barycentric {{} osculating} semi-major axis and eccentricity of the resonant test particles over their 10 Myr evolution, as found in the N-body numerical simulations.}
 \label{f:f6}
\end{figure}

\begin{figure}
 \centering
 \includegraphics[width=250px]{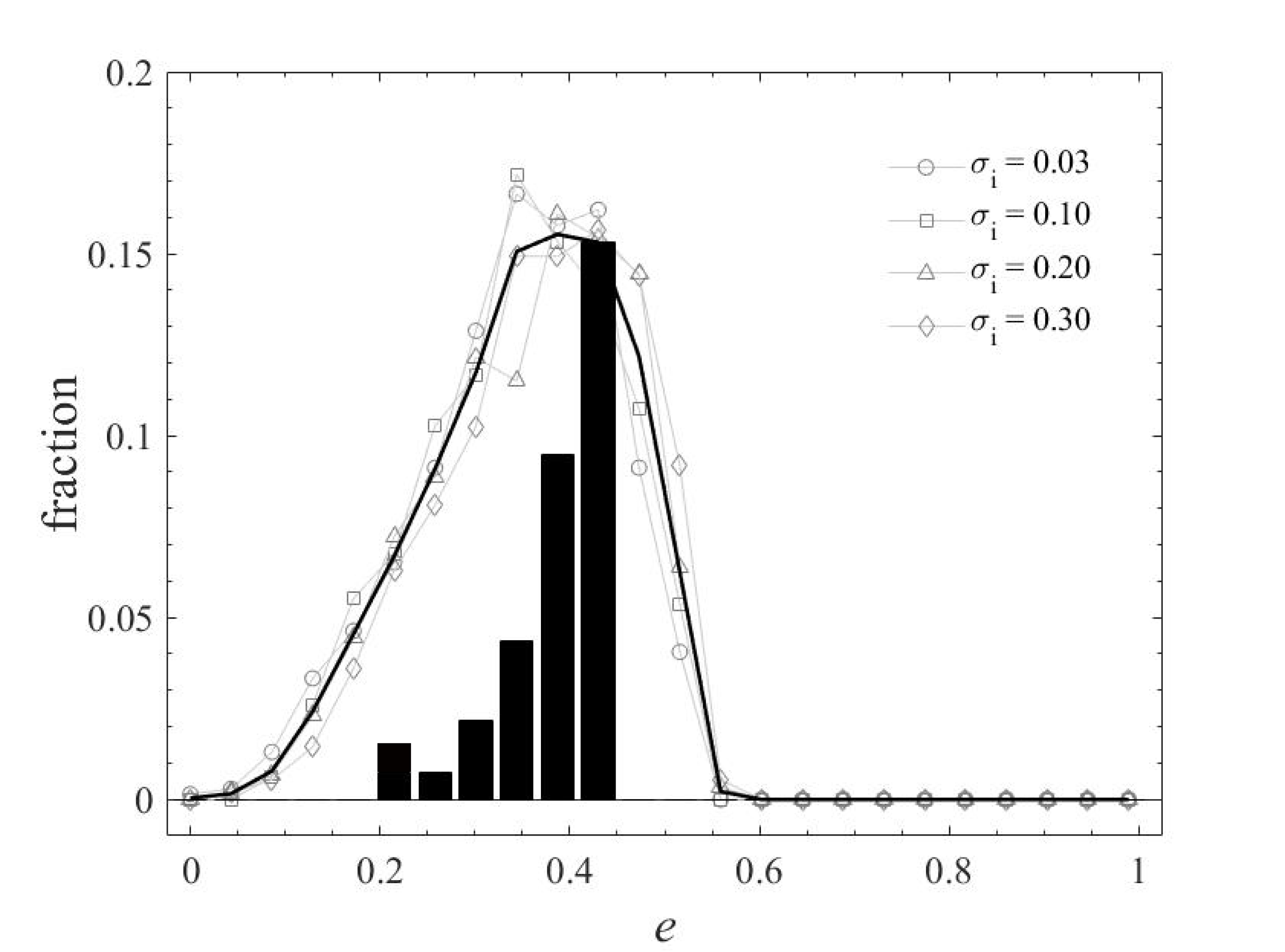}
 \caption{The {{} gray and black curves indicate the distribution of the barycentric osculating eccentricity of the} resonant particles surviving at the end of the N-body numerical simulations. The point-styles with circles, squares, triangles and diamonds indicate the distribution obtained in Sim-1, Sim-2, Sim-3 and Sim-4, respectively. The solid black line is the average of the four distributions. The black histogram indicates the distribution of the observational sample, whose peak is scaled to the peak of the solid black line.}
 \label{f:f7}
\end{figure}

\begin{figure}
 \centering
 \includegraphics[width=250px]{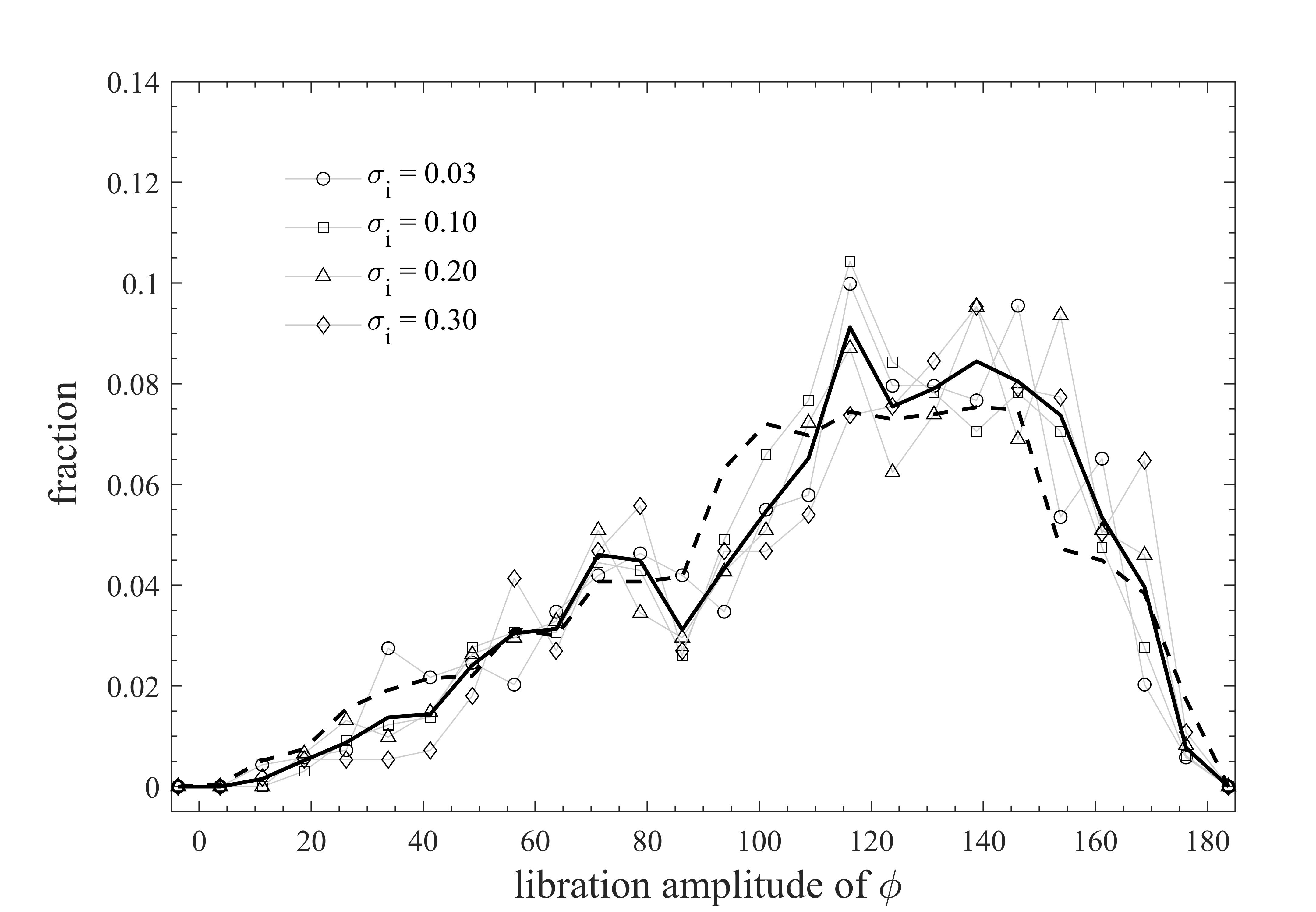}
 \caption{Distribution of the libration amplitude of the resonant angle, $\phi$ (Eq.~\ref{e:phi}) of the resonant particles from four simulations. The point-styles with circles, squares, triangles and diamonds indicate the distribution obtained in Sim-1, Sim-2, Sim-3 and Sim-4, respectively.  The solid black line is the average of the four distributions. 
 }
 \label{f:f8}
\end{figure}


In the previous section, we identified the boundaries of the stable zones of Neptune's 5:2 MMR with the simplified planar circular restricted three body model of the Sun, Neptune and test particle KBOs.  For a more complete picture of the long term stability in this resonance, we investigated the behavior of test particles in the neighborhood of this resonance in three dimensions with full N-body numerical simulations under the gravitational perturbations of all four giant planets, Jupiter--Neptune.
We carried out four separate simulations, each with 6000 test particles with initial conditions uniformly distributed in a grid in $(a,e)$, with $a$ in the range $1.82a_{\rm{N}}$ to $1.86a_{\rm{N}}$ (where $a_N$ is Neptune's initial semi-major axis), and $e$ is in the range 0 to 1.  In Sim-1, Sim-2, Sim-3 and Sim-4, the initial inclinations, $i$ (relative to the solar system's invariable plane), were chosen from a Rayleigh distribution\footnote{{}The probability density function of the Rayleigh distribution is $f(x) = (x/\sigma^2)\exp(-x^2/2\sigma^2)$ where $\sigma$ is the scale parameter; the mode of this distribution is $\sigma$, its mean is $(\pi/2)^{1\over2}\sigma$, and its variance is $(4-\pi)\sigma^2/2$.} of $\sin i$ with {{} scale} parameter $\sigma_i=0.03, 0.1, 0.2$ and 0.3 respectively. The remaining orbital parameters (the mean anomaly, the argument of perihelion and the longitude of ascending node) were uniformly distributed  in the range 0 to $2\pi$.  We integrated for $10^7$~years, with output every $10^3$ years.  We used the second order mixed variable symplectic integrator \citep{Wisdom:1991}, with a step size of 0.1 year.  The integration self-consistently accounts for the orbital perturbations of the four giant planets on each other and on the test particles.  We discarded any particles that approached a planet within a Hill sphere distance, or whose heliocentric distance decreased below 5 au or exceeded 1000 au.  The surviving particles were examined for libration of the resonant angle, $\phi=5\lambda-2\lambda_{N}-3\varpi$.  We identified the resonant particles as those exhibiting librations of $\phi$ about either a center at $180^\circ$ (Zone 1-1 or Zone 1-2) or a center at $0^\circ$ (Zone II), with a maximum libration amplitude not exceeding $175^\circ$.

At the end of the 10 megayear integrations, we found approximately 10\% of the initial population of test particles librating in Zone I-1, none in Zones I-2 and II. We found little difference in the $(a,e)$ distribution of the survivors in the four simulations.  A scatter plot of the time history over 10 Myr of the barycentric osculating semi-major axis and eccentricity of all the librating particles is shown in Figure~\ref{f:f6}.  
We observe that surviving librating particles are bounded in the $(a,e)$ plane with boundaries matching quite well with the boundaries of Zone I-1 from the simplified model investigated in Section 2, but limited to eccentricities below $\sim0.53$ (corresponding to perihelion distance $\sim 26$~au); higher eccentricity test particle orbits did not survive in any resonant zone to the end of the 10 megayears.   

We note that the choice of barycentric orbital elements is preferable for illustrating resonant structure in the outer solar system.  In heliocentric coordinates, the osculating semi-major axis of a KBO has short period fluctuations due to the motion of the Sun. These fluctuations are dominated by a $\sim12$ year period (due to Jupiter's effect on the Sun's motion) and have amplitude of $\sim0.5$~au (for semi-major axes near the 5:2 MMR), roughly half as large as the 5:2 resonance width. The resonance boundaries become clear only when this short-period variation is removed by considering barycentric osculating elements.

For the librating resonant particles in each of the four simulations, the distribution of eccentricities and of the libration amplitudes of the resonant angle, $\phi$, is shown in Figure~\ref{f:f7} and Figure~\ref{f:f8}, respectively.  We observe that the initial distribution of inclination has little effect on the eccentricity distribution.  Similarly the distribution of the libration amplitude of $\phi$ is not strongly sensitive to the initial inclination distribution.  The eccentricity distribution is roughly triangular in shape, it increases steeply with eccentricity from $e\approx0.1$ up to the peak near 0.4, then drops steeply between the peak and a maximum near 0.53. The median is near $e\approx0.35$; approximately 20\% fraction is found between eccentricity 0.45 and 0.53.  The libration amplitude distribution has a broad peak near 100--150$^\circ$; there is a hint of fine structure in this distribution, with a local peak near $70^\circ$. The fine structure is possibly due to secondary resonances arising from commensurability of the libration period with other long period perturbations induced by either the eccentricity-inclination coupling or by the perturbations of the other giant planets.  Analogous dynamical features have been identified within other Neptune MMRs \citep[e.g.,][]{Milani:1989,Morbidelli:1995,Morbidelli:1997, Kortenkamp:2004,Tiscareno:2009}.

\section{Comparison with observations}\label{s:observations}

The first KBOs identified as inhabiting Neptune's 5:2 MMR were recognized by \cite{Chiang:2003}.  Additional 5:2 resonators have been identified by \cite{Lykawka:2007b}, \cite{Gladman:2008}, \cite{Gladman:2012} and \cite{Volk:2016}.
To identify resonant objects from observational data requires two steps: (a) astrometric data are fit to a Keplerian orbit model to derive best-fit orbital parameters and their uncertainties; (b) the KBO's orbit is integrated forward in time (with a numerical integrator) in a numerical model of the solar system to diagnose whether the resonant angle, $\phi$, persists in a libration state.

We downloaded the available astrometric data in the Minor Planet Center for those objects listed with heliocentric semi-major axes in the range $54-57$~au. (This range generously covers the width of the 5:2 MMR determined in Sections 2 and 3 above, after accounting for the difference between heliocentric and barycentric osculating semi-major axis.) For those objects observed over more than one opposition, we used the \citet{Bernstein:2000} orbit fitting code to find the best-fit orbit for each object;  objects with shorter observational arcs have semi-major axis uncertainties too large to reliably classify. Starting from the orbit-fit epoch, these orbits were integrated forward in time for 10 Myr as massless test particles with the four giant planets and the Sun as massive perturbers. Following the approach of \citet{Gladman:2008}, we determined which objects librate within the 5:2 MMR for at least half of the 10 Myr duration. We identified 46 resonant objects, all of which are found to be librating in the resonance Zone I-1.  In Table~\ref{t:52}, we list these objects, along with their barycentric osculating elements (semi-major axis, eccentricity, and inclination), as well as the uncertainty, $\sigma_a$, in the semi-major axis determined from the covariance matrix of the \citet{Bernstein:2000} orbit-fit. {{} For each object, the orbit fit was calculated at its discovery epoch.  The discovery epochs of this observational sample span approximately 20 years. The listed barycentric osculating orbital elements do not significantly change over this small time span.}

The 5:2 resonators in the observational sample are indicated by the black dots in Figure~\ref{f:f3}, where we make a scatter plot of their best-fit barycentric semi-major axes and eccentricities in the $(a,e)$ plane. We observe that the observational sample is concentrated in the region of the maximum width of the libration Zone I-1 in the $(a,e)$ plane.

\begin{table}
\caption{The 5:2 resonators identified from astrometric data retrieved from the Minor Planet Center. The listed orbital elements are barycentric osculating elements {{} at the object's discovery epoch} (referred to the J2000 ecliptic-equinox), and $\sigma_a$ is the uncertainty in the semi-major axis determined from the covariance matrix of the \citet{Bernstein:2000} orbit-fit.
}
\begin{center}
\begin{tabular}{l  l l l l}
MPC designation & $a$ (au) & $\sigma_a$ (au) & $e$ & $i$ (deg) \\
\hline 
26375 & 55.405 & 0.002 & 0.418 & 7.62 \\
38084 & 55.60 & 0.01 & 0.414 & 13.151 \\
60621 & 55.29 & 0.03 & 0.402 & 5.869 \\
69988 & 55.18 & 0.01 & 0.429 & 9.457 \\
84522 & 55.288 & 0.004 & 0.293 & 35.038 \\
119068 & 55.09 & 0.01 & 0.357 & 12.905 \\
135571 & 55.29 & 0.01 & 0.351 & 14.684 \\
143707 & 55.55 & 0.01 & 0.415 & 7.537 \\
471151 & 55.314 & 0.004 & 0.423 & 10.726 \\
471172 & 55.222 & 0.004 & 0.43 & 3.126 \\
472235 & 55.424 & 0.004 & 0.409 & 0.807 \\
495603 & 55.282 & 0.006 & 0.25 & 26.729 \\
1999 RU214 & 55.70 & 0.03 & 0.341 & 4.26 \\
2000 SR331 & 55.382 & 0.008 & 0.438 & 4.257 \\
2001 XQ254 & 55.37 & 0.01 & 0.439 & 7.109 \\
2002 GP32 & 55.387 & 0.007 & 0.422 & 1.559 \\
2004 EG96 & 55.527 & 0.007 & 0.423 & 16.213 \\
2004 HO79 & 55.21 & 0.01 & 0.412 & 5.624 \\
2004 KZ18 & 55.42 & 0.02 & 0.382 & 22.646 \\
2004 TT357 & 55.34 & 0.02 & 0.431 & 8.981 \\
2005 SD278 & 55.50 & 0.01 & 0.282 & 17.851 \\
2005 XN113 & 55.360 & 0.006 & 0.405 & 3.383 \\
2007 LG38 & 55.457 & 0.006 & 0.434 & 32.579 \\
2009 YG19 & 55.666 & 0.006 & 0.408 & 5.154 \\
2011 UT411 & 55.7 & 0.3 & 0.406 & 6.42 \\
2012 UJ177 & 55.20 & 0.07 & 0.433 & 15.632 \\
2012 UD178 & 55.27 & 0.02 & 0.418 & 8.694 \\
2013 GS136 & 55.63 & 0.03 & 0.385 & 6.978 \\
2013 GY136 & 55.55 & 0.03 & 0.414 & 10.877 \\
2013 JF64 & 55.42 & 0.01 & 0.45 & 8.785 \\
2013 JK64 & 55.250 & 0.009 & 0.408 & 11.077 \\
2013 RQ98 & 55.7 & 0.1 & 0.434 & 37.69 \\
2013 UD17 & 55.46 & 0.03 & 0.396 & 26.372 \\
2013 UO17 & 55.006 & 0.004 & 0.389 & 3.833 \\
2014 FM72 & 55.4 & 1.6 & 0.389 & 4.716 \\
2014 HA200 & 55.70 & 0.02 & 0.342 & 10.385 \\
2014 JX80 & 55.52 & 0.02 & 0.355 & 28.868 \\
2014 KB102 & 55.47 & 0.02 & 0.409 & 11.122 \\
2014 US229 & 55.261 & 0.003 & 0.398 & 3.902 \\
2014 UA230 & 55.33 & 0.01 & 0.232 & 12.705 \\
2014 UO231 & 55.415 & 0.009 & 0.237 & 23.156 \\
2014 WS510 & 55.261 & 0.009 & 0.375 & 8.907 \\
2014 YL50 & 55.2 & 0.3 & 0.318 & 29.145 \\
2015 BC519 & 55.43 & 0.04 & 0.417 & 1.723 \\
2015 BD519 & 55.21 & 0.06 & 0.345 & 10.368 \\
2015 PD312 & 55.39 & 0.02 & 0.374 & 23.06 \\
\end{tabular}
\end{center}
\label{t:52}
\end{table}%

\begin{figure}
 \centering
 \includegraphics[width=85mm]{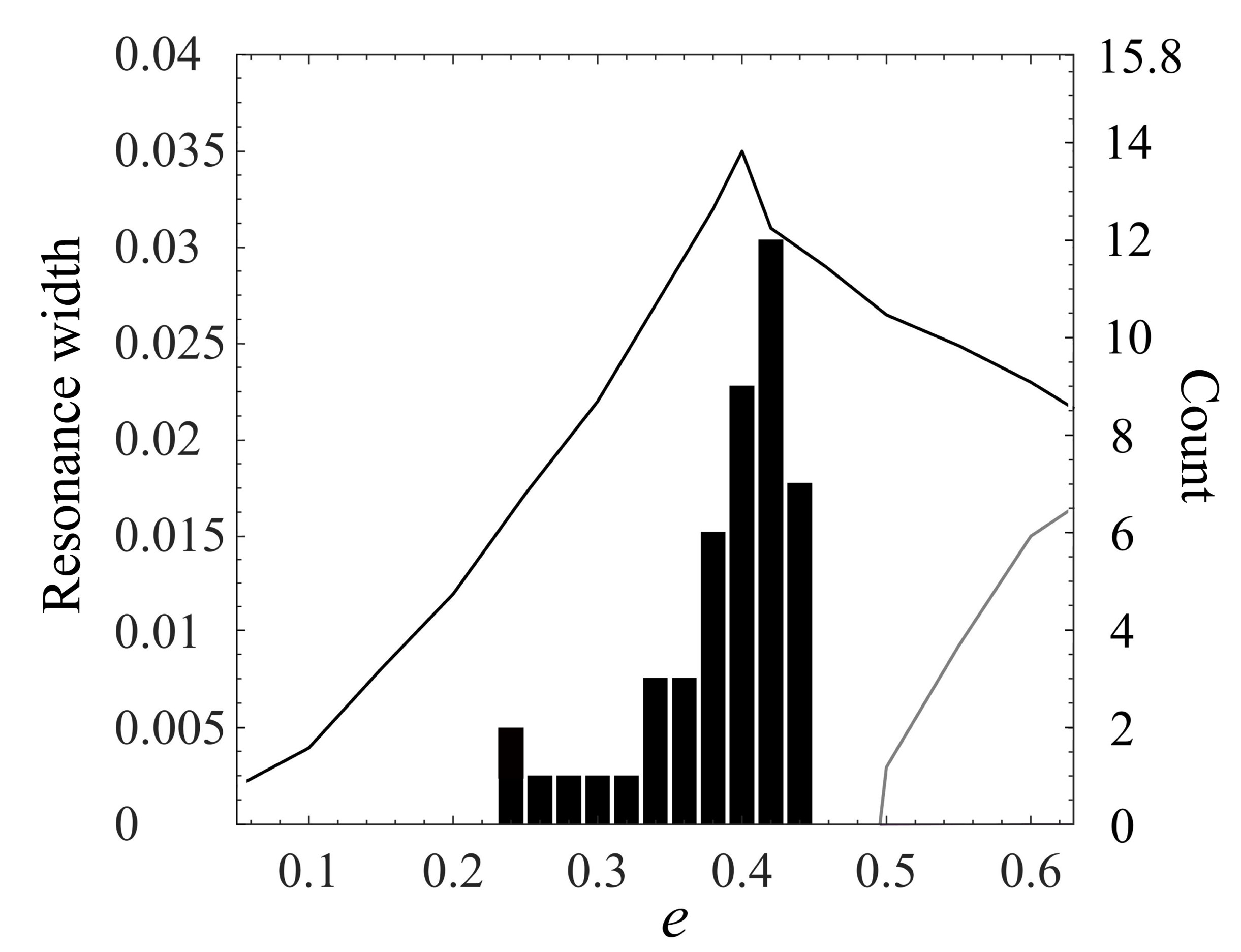}
\caption{The black histogram shows the distribution of the {{} barycentric osculating} eccentricity  of the detected objects in Neptune's 5:2 exterior MMR; the vertical scale for this histogram is indicated on the right side of the figure. The resonance width, $\Delta a/a_N$, as a function of the test particle's eccentricity is shown with the black curve (for Zone I-1) and with the gray curve (for Zone II); the vertical scale is indicated on the left side of the figure.
}
 \label{f:f9}
\end{figure}

In Figure~\ref{f:f9}, we plot the histogram of the eccentricities of the observational sample. We also plot in this figure the width, $\Delta a$, of the resonant zones I-1 and II as a function of eccentricity (as computed with the circular planar restricted three body model, Section~\ref{s:pcr3bd}).
We see that the eccentricity distribution of the observational sample drops off more steeply at eccentricities above and below $e\approx0.4$ compared to the corresponding decrease in the width of the resonance Zone I-1.  In Figure~\ref{f:f7}, we compare the eccentricity histogram of the observational sample with the eccentricity distributions obtained from the four numerical simulations of the three-dimensional model including all four giant planets' perturbations (Section~\ref{s:numerical-sims}).  Again, we see that the observations have a steeper drop-off away from the peak at $e\approx0.4$ than found in the modeled eccentricity distribution.

For the higher eccentricities, we can understand the lack of observed resonant KBOs above $e\approx 0.53$, based on their lack of long term stability under the combined perturbations from all four giant planets, as illustrated in Figure~\ref{f:f6}.  However, we note that the observational sample has a maximum eccentricity near $\sim0.45$, whereas in the simulations we find approximately 15\% of the surviving resonators having larger eccentricities. Is the absence of resonant objects with eccentricities between 0.45 and 0.53 of significance? The explanation may lie in the limited duration --10 megayears-- of our numerical simulations: over the $\sim4.5$~gigayear age of the solar system, planetary perturbations may render unstable the resonant orbits in this eccentricity range; a definitive answer to this question is left for a future study.

For the lower eccentricities, the sharp drop-off in the population of observed objects is likely owed to observational bias, as lower eccentricities correspond to higher perihelion distances, rendering such KBOs fainter and more difficult to detect.  (For example, for the same absolute magnitude, an object in the 5:2 MMR with $e=0.25$ will be about one magnitude fainter at perihelion than an object with $e=0.4$.)  A quantitative assessment of the observational bias is not feasible because this sample of objects were observed by many different observers under different observational conditions.  However, we can compare the eccentricity distribution found for the resonant survivors in the 10 Myr N-body numerical simulations (shown with the black curve in Figure~\ref{f:f7}) with the results of \citet{Volk:2016} who modeled the intrinsic eccentricity distribution of a small sample of the 5:2 resonators discovered in the first quarter of the well-characterized Outer Solar System Origins Survey \citep[OSSOS;][]{Bannister:2016}. In their model, they used a Gaussian function for the eccentricity distribution and determined a range of mean eccentricity and standard deviation consistent with the observations. The eccentricity distribution we find for the simulated resonant survivors can be roughly approximated with a Gaussian function of mean $\bar e\simeq0.38$ and a standard deviation $\sigma_e\simeq0.1$; these values lie within the range of models found to be acceptable by \citet{Volk:2016} for the OSSOS data. Future, larger observational samples of the 5:2 resonators will be useful to further constrain their eccentricity distribution. 

We note the lack of any Zone II objects in the observational sample. This is consistent with the absence of any Zone II survivors in our 10 megayear numerical simulations (see Figure~\ref{f:f6}).  However, we found one candidate amongst the observed KBOs, the object with MPC designation 2013 UR$_{15}$, which appears to be temporarily resident in Zone II.  We discuss this object in the Appendix.

The libration amplitude distribution obtained in our numerical simulations is shown in Figure~\ref{f:f8}.  In these simulations, we used initial conditions that uniformly populated the $(a,e)$ range of the 5:2 MMR, orbital inclinations were taken to have a Rayleigh distribution about the solar system's invariable plane, and other angular elements were chosen randomly. Proposed models of the outer solar system's dynamical history can be expected to produce resonant KBO populations whose libration amplitude distributions may be distinct \citep{Chiang:2003,Levison:2008}.  The distribution of the libration amplitude of resonant populations may therefore offer constraints on their origin mechanisms.  
With current data, the orbit-fits for many of the observed 5:2 resonators are not sufficiently accurate to securely determine libration amplitudes; 
longer observational arcs are needed to reduce the uncertainties in the measured libration amplitudes to useable values. 
Moreover, the observational biases in libration amplitude depend on the design of surveys in which the objects were detected \citep[see, e.g.,][]{Gladman:2012,Volk:2016}.
As future surveys increase the observational arc lengths as well as the sample size of 5:2 resonators with characterized detection biases, better estimates of the eccentricity and libration amplitude distributions will allow more rigorous constraints on their origins.

\section{Summary and Conclusions}
 
In the present work, we have investigated in some detail the dynamical structure of Neptune's 5:2 MMR in order to understand some of the puzzles presented by the observed population of KBOs in this resonance.
As a first step, we made use of Poincar\'e sections of the circular planar restricted three body model.  This is the simplest dynamical model to study the widths and strengths of Neptune's MMRs in the Kuiper belt.  For a range of values of the Jacobi integral, we explored the resonant structure in the $(a,\psi)$ surfaces-of-section, where $a$ is the test particle's osculating semi-major axis and $\psi$ is the angular separation of Neptune from the test particle at pericenter.  In these surfaces-of-section, the stable resonant zones are readily identified as a pair of islands containing smooth closed curves surrounding the exact resonant locations, corresponding to orbits librating about the exact resonance.

With increasing values of the particle eccentricity, from $\sim0$ to $\sim1$, we found that the structure of the resonance in the Poincar\'e sections shows five transitions (illustrated in Figure~\ref{f:f2}). 
These transitions observed in the Poincar\'e sections are correlated with the eccentricity-dependent changing shape and geometry of the trace of the particle's resonant trajectory in the rotating frame (rotating with the angular frequency of the primaries about an axis normal to their common orbital plane and centered at their barycenter).
The centers and widths in $\psi$ of the resonant islands are related to the lengths of the arcs of the circle of radius $1-\mu$ which lie interior and exterior to the perihelion lobes of the particle's resonant trajectory in the rotating frame.
We identify three distinct resonant zones:
Zone I-1 exists in the eccentricity range $e<e_{\rm{c2}}\simeq0.872$ and has libration of the resonant angle $\phi$ centered at $180^\circ$;  its maximum width in semi-major axis occurs at eccentricity $e_{\rm{m1}}\simeq0.402$.
Zone II exists in the eccentricity range $e>e_{\rm{c1}}\simeq0.473$ and has libration of the resonant angle $\phi$ centered at $0^\circ$; its maximum width in semi-major axis occurs at eccentricity $e_{\rm{m2}}\simeq0.874$.
Zone I-2 exists in the eccentricity range $e>e_{\rm{c3}}\simeq0.909$ and has libration of the resonant angle $\phi$ centered at $180^\circ$;  this is a small zone at nearly parabolic eccentricities.
The appearance of Zone I-2 arises from the self-intersection of the perihelion lobes when the eccentricity exceeds $e_{\rm{c3}}$.

By examination of many Poincar\'e sections, we measured the boundaries of the resonant islands of Zone I-1, Zone I-2 and Zone II, and thereby obtained an accurate map of the boundaries of the stable libration zone of Neptune's 5:2 MMR in the $(a,e)$ plane for the full range of eccentricities, (0--1); this is shown in Fig.~\ref{f:f3}.
{{} In Figure~\ref{f:MMRcompare}, we compare the size of the stable 5:2 MMR in the $(a,e)$ plane with the sizes of the 3:2 and the 2:1 MMRs. This comparison shows that the size of Neptune's 5:2 MMR (Zone I-1) is rather similar to the sizes of 3:2 and 2:1 MMRs.}
A noteworthy coincidence is that the maximum widths of these three different MMRs all occur at eccentricities corresponding to a similar perihelion distance, $\sim33$ au.  

Going beyond the simplified planar circular restricted three body model, we carried out three-dimensional N-body numerical simulations with all four giant planets and thousands of test particle KBOs, integrating their orbits for 10 megayears; these simulations include a wide range of orbital inclinations of the KBOs relative to the solar system's invariable plane.  From these simulations, we find that the boundaries of the stable zone of the 5:2 MMR in the semimajor axis--eccentricity plane are very similar to those found with the simplified circular planar restricted three body model of the Sun-Neptune-KBO {{} (Figure~\ref{f:f6})}, with the caveat that orbits of eccentricity above $\sim0.53$ are long term unstable; such orbits, which have perihelion distance less than $\sim$~26 AU, are phase-protected from close encounters with Neptune but not from destabilizing encounters with Uranus. Additionally, the numerical simulations show that the long term stability of KBOs in Neptune's 5:2 MMR is not strongly sensitive to KBO inclination.

With current astrometric data of the observed KBOs, we computed the best-fit barycentric orbital parameters and their uncertainties for those KBOs having orbits near Neptune's 5:2 MMR.  We numerically integrated the orbits of these objects for 10 megayears and identified 46 objects which exhibit persistent libration of their 5:2 resonant angle (centered at $\phi=180^\circ$) for at least 5 megayears.  Table 1 lists this sample, Figure~\ref{f:f3} shows a scatter plot of their best-fit {{} barycentric osculating} semi-major axes and eccentricities, Figure~\ref{f:f7} compares their eccentricity distribution with that found in the three-dimensional N-body numerical simulations, and Figure~\ref{f:f9} compares their distribution with the resonance width. (One notable candidate for the 5:2 resonance not listed in Table 1 is 2013 UR$_{15}$; this object is likely a temporary resident of resonance Zone II and is described in the Appendix.)

{{} Considering these results, the properties of the observed 5:2 resonators -- their concentration near eccentricity $\sim0.4$ and their intrinsic population estimates being comparable to those of the Plutinos and Twotinos -- do not appear as puzzling as at first sight.  The concentration of their eccentricities near 0.4 is congruent with the resonance width maximum near this value of eccentricity (Figure~\ref{f:f9}); at lower eccentricities the resonance width decreases sharply, and at higher eccentricities the particles are long term unstable due to close encounters with Uranus.  The size of the long term stable 5:2 resonance zone in the $(a,e)$ plane is similar to the corresponding sizes of the stable libration zones of the 3:2 and 2:1 MMRs (Figure~\ref{f:MMRcompare}). This similarity is broadly congruent with the similarities of the estimated intrinsic populations of the MMRs.} 

A remaining question is the origin of the 5:2 resonators: what mechanism excited the eccentricities and inclinations of this population from their presumed formation within a dynamically cold planetesimal disk? Proposed mechanisms include gravitational scattering and/or convergent resonance sweeping with an outwardly migrating Neptune.  {{} \cite{Chiang:2003} found that the resonance sweeping scenario of \cite{Malhotra:1993,Malhotra:1995} could yield capture efficiencies in the 5:2 MMR approaching those of the 2:1 MMR if Neptune migrated smoothly into a pre-stirred planetesimal disk, a result confirmed by~\cite{Hahn:2005}. Alternatively, gravitational scattering during an instability-driven migration of the giant planets could also lead to capture of eccentric and inclined resonant populations, although published models generally predict a significantly less populated 5:2 MMR than the 2:1 and 3:2 MMRs~\citep{Levison:2008,Gladman:2012}.} With more accurate orbital parameters and larger observational samples, {{} more accurate quantitative comparisons of the population ratios and} the detailed distribution of the eccentricity, inclination and libration amplitude of the resonant angle may help to diagnose the origin mechanisms.

In the present work, we have discussed only the 5:2 resonators.  {{} Observational data indicate that some other higher order Neptune's MMRs -- such as the 5:3 and the 7:4 -- host significant populations of KBOs  \citep{Gladman:2012,Adams:2014,Volk:2016}}. In a future study, it would be meaningful to investigate these MMRs with similar methods as those used in the present work to more accurately assess the dynamical structures and widths/strengths of those MMRs and make comparisons of the abundances in the various MMRs.

\vglue0.2truein

{{} We thank an anonymous referee for comments that improved the paper.} RM and KV acknowledge funding  from NASA (grant NNX14AG93G). 
LL acknowledges funding from National Natural Science Foundation of China (11572166) and China Scholarship Council.

\appendix

\begin{figure}[h!]
 \centering
 \includegraphics[width=240px]{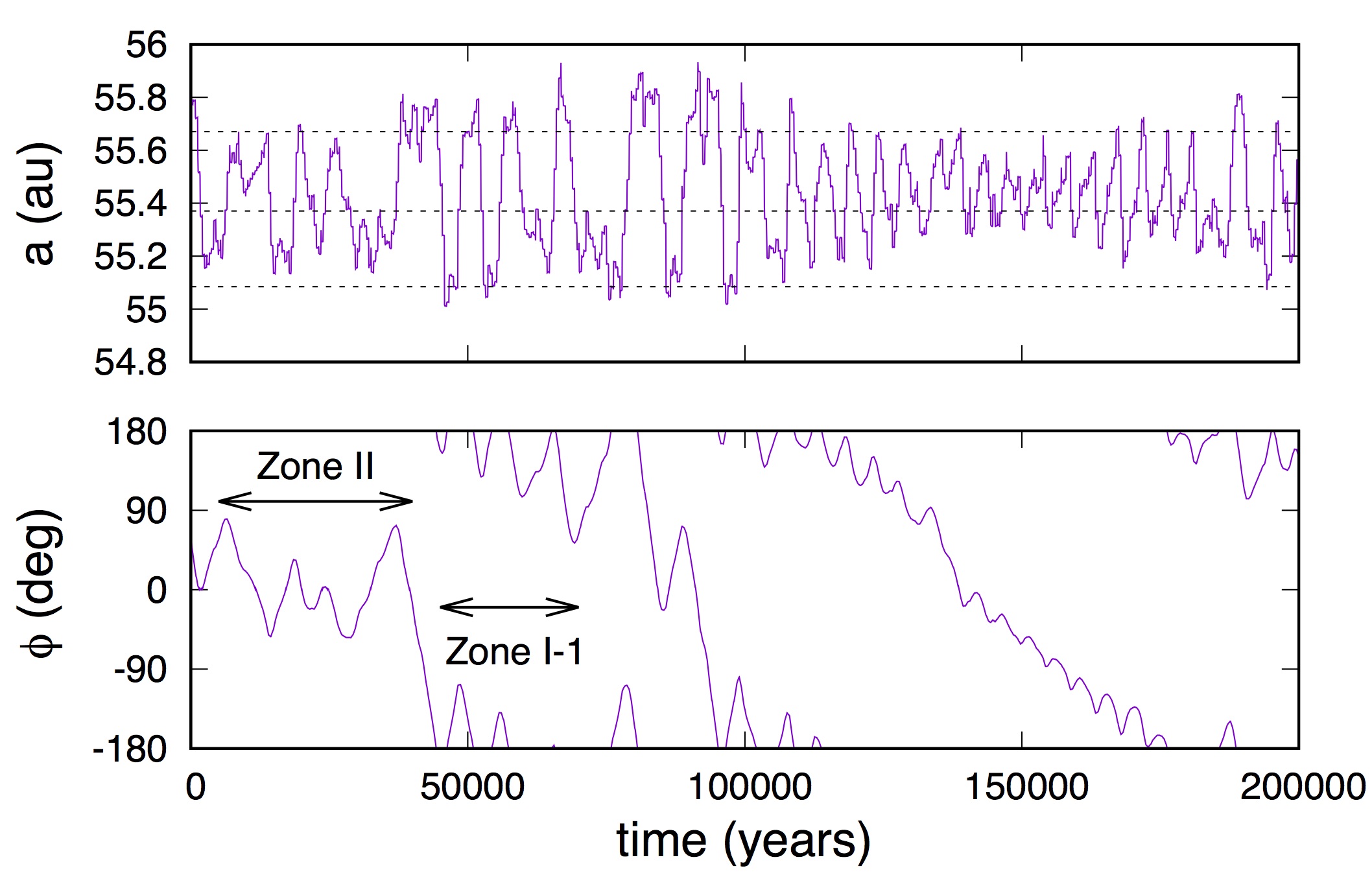}
 \caption{Time variation of the semi-major axis (top) and resonant angle $\phi$ (bottom) for a clone of KBO 2013 UR$_{15}$. The dashed lines in the top panel indicate the center of the resonance and the maximum range of semi-major axis of the libration islands of Zone II, at $e=0.72$.}
 \label{f:obs-zone2}
\end{figure}

In Section~\ref{s:observations}, we noted the lack of any Zone II objects in the observational sample, based on the criterion that the resonant angle $\phi$ exhibits persistent libration for at least 5 megayears.  A close examination of the 10 megayears numerical integrations of the observed KBOs near Neptune's 5:2 MMR revealed one candidate for temporary residence in the 5:2 MMR's Zone II: the KBO designated as 2013 UR$_{15}$.  This object, {{} discovered in the Outer Solar Systems Origins Survey \citep{Bannister:2018},} has a current semi-major axis very close to the 5:2 MMR. Its eccentricity, $e=0.72$, corresponds to a perihelion distance $\sim15.5$~au, interior to Uranus' orbit.  In the numerical integrations, we find that on megayear timescales, this object scatters away from the resonance due to close encounters with the planets. However, on timescales of tens of thousands of years, this object is likely to exhibit librations of the resonant angle $\phi$ centered at $\phi=0$, in Zone II.  Figure~\ref{f:obs-zone2} shows the semi-major axis and resonant angle evolution of a clone of this object for a duration of 200 thousand years, with initial conditions at the present epoch.  We integrated 250 clones of this object's orbit, varying the initial orbital parameters within their uncertainties as determined by the covariance matrix of the best-fit orbit from \cite{Bernstein:2000}'s orbit fitting procedure. We find that most of the clones show behavior similar to Figure~\ref{f:obs-zone2}: libration in Zone II for a few cycles, then intermittent Zone I-1 librations and circulations of $\phi$. The semi-major axis variations during the period of libration are similar to the resonant width found in Figure~\ref{f:f3} for 2013 UR$_{15}$'s eccentricity.
This indicates that although resonant orbits in Zone II are long term unstable, short term ``stickiness" in this high eccentricity resonance zone remains a real possibility for the ``scattered" and ``scattering" populations of KBOs.  We suggest that orbit-fitting efforts for resonant KBOs should not neglect the possibility of such high eccentricity resonant orbits.  This means search algorithms for identifying resonant KBOs should include the possibility of resonant angle libration centers different than the usual center at $180^\circ$.

\vglue0.3truein


\end{document}